\documentclass[12pt,onecolumn,draftcls]{IEEEtran} 
\usepackage[pdftex]{graphicx}
\usepackage{xcolor}
\usepackage{epstopdf}
\usepackage{psfrag,graphicx}
\usepackage{verbatim}
\usepackage{amsfonts}
\usepackage{algorithmic}
\usepackage[font=small]{caption}
\usepackage{amssymb}
\usepackage{amsmath}
\usepackage{amsthm}
\usepackage{subfigure}
\usepackage{epsfig}
\usepackage{subfig}
\usepackage{varioref}
\usepackage{blkarray}
\usepackage{graphicx} 
\usepackage{color}
\usepackage{enumerate}
\usepackage[ruled,vlined]{algorithm2e}

\theoremstyle{newstyle}
\newtheorem{theorem}{Theorem}

\newtheorem{lemma}[theorem]{Lemma}

\newtheorem{claim}[theorem]{Claim}

\newtheorem{definition}{Definition}

\newcommand{\expectation}{\operatorname{E}}
\newcommand{\ones}{\mathbf{1}}

\newcommand{\re}{\operatorname{Re}}
\newcommand{\im}{\operatorname{Im}}

\newcommand{\lb}{\left(}
\newcommand{\rb}{\right)}
\newcommand{\lsb}{\left[}
\newcommand{\rsb}{\right]}

\newcommand{\beq}{\begin{equation}}
\newcommand{\eeq}{\end{equation}}
\newcommand{\bea}{\begin{array}}
\newcommand{\ena}{\end{array}}
\newcommand{\bds}{\begin {itemize}}
\newcommand{\eds}{\end {itemize}}
\newcommand{\bdf}{\begin{definition}}
\newcommand{\blm}{\begin{lemma}}
\newcommand{\edf}{\end{definition}}
\newcommand{\elm}{\end{lemma}}
\newcommand{\bthm}{\begin{theorem}}
\newcommand{\ethm}{\end{theorem}}
\newcommand{\bprp}{\begin{prop}}
\newcommand{\eprp}{\end{prop}}
\newcommand{\bcl}{\begin{claim}}
\newcommand{\ecl}{\end{claim}}
\newcommand{\bcr}{\begin{coro}}
\newcommand{\ecr}{\end{coro}}
\newcommand{\bquest}{\begin{question}}
\newcommand{\equest}{\end{question}}

\newcommand{\yvec}{{\bf{y}}}

\newcommand{\vvec}{{\bf{v}}}

\newcommand{\hvec}{{\bf{h}}}

\newcommand{\Hmat}{{\bf{H}}}

\newcommand{\Rmat}{{\bf{R}}}

\newcommand{\be}{\begin{equation}}
\newcommand{\ee}{\end{equation}}
\newcommand{\beqna}{\begin{eqnarray}}
\newcommand{\eeqna}{\end{eqnarray}}

\newcommand{\symbolconstellation}{\mathcal{P}}

\newcommand{\argmax}{\operatorname{argmax}}

\begin{document}

\sloppy

\title{Energy-based Modulation for \\ Noncoherent Massive SIMO
  Systems}

\author{ Alexandros Manolakos, Mainak Chowdhury and Andrea Goldsmith
  \IEEEmembership{Fellow, IEEE} \thanks{ The authors are with the
    Department of Electrical Engineering, Stanford University,
    Stanford, CA - 94305.  Questions or comments can be addressed to
    \{amanolak, mainakch, andreag\}@stanford.edu.  Parts of this work
    were presented at IEEE Globecom, 2014.  This work is supported by
    the Alcatel-Lucent Corporation Stanford Graduate Fellowship, the
    3Com Stanford Graduate Fellowship, an A.G. Leventis Foundation
    scholarship, NSF grant 1320628, ONR grant N000141210063, and a
    research grant from CableLabs. }}
\maketitle

%% Abstract:
%%
\begin{abstract}
  An uplink system with a single antenna transmitter and a single
  receiver with a large number of antennas is considered.  We propose
  an energy-detection-based single-shot noncoherent communication
  scheme which does not use the instantaneous channel state
  information (CSI), but rather only the knowledge of the channel
  statistics. The suggested system uses a transmitter that modulates
  information on the power of the symbols, and a receiver which
  measures only the average energy across the antennas. We propose
  constellation designs which are asymptotically optimal with respect
  to symbol error rate (SER) with an increasing number of antennas,
  for any finite signal to noise ratio (SNR) at the receiver, under
  different assumptions on the availability of CSI statistics (exact
  channel fading distribution or the first few moments of the channel
  fading distribution).  We also consider the case of imperfect
  knowledge of the channel statistics and describe in detail the case
  when there is a bounded uncertainty on the moments of the fading
  distribution.  We present numerical results on the SER performance
  achieved by these designs in typical scenarios and find that they
  may outperform existing noncoherent constellations, e.g.,
  conventional Amplitude Shift Keying (ASK), and pilot-based schemes,
  e.g., Pulse Amplitude Modulation (PAM).  We also observe that an
  optimized constellation for a specific channel distribution makes it
  very sensitive to uncertainties in the channel statistics.  In
  particular, constellation designs based on optimistic channel
  conditions could lead to significant performance degradation in
  terms of the achieved symbol error rates.
\end{abstract}

\section{Introduction}

Accurate channel state acquisition is critical to realizing many of
the diversity and beamforming benefits in existing high speed
communication systems.  With an increasing number of antennas and
operating bandwidths, the complexity and overhead of channel state
acquisition grows proportionally.  However, in a recent result
\cite{mainak2014, mainak2014journal} for a narrowband massive SIMO
system with independent and identically distributed channel
coefficients at each of the receiver antennas, it was shown that one
can achieve a scaling law in the number of receive antennas which is
no different from that with perfect channel state information (CSI).
The achievability uses only an average energy measurement across all
the receiver antennas.  In another recent work \cite{mainakITW2015},
it was shown that employing a similar energy-based scheme in a
noncoherent massive wideband SIMO system leads to optimal capacity
scaling when both the number of receive antennas and the bandwidth go
to infinity jointly.  These asymptotic characterizations suggest that
when the number of antennas is large, the requirements on the channel
state information acquisition may not be very
stringent. % However, even
% though all these results are surprising, they are all asymptotic; to
% achieve reasonable bit error rates (BERs) according to the proposed
% constellation designs shown in these works, one would need on the
% order of $1000$ receive antennas.

Motivated by these asymptotic results, in this work, we consider a
finite number of receive antennas in a massive SIMO system and address
the problem of optimizing the transmit constellation points based on
an asymptotically tight and analytically tractable upper bound on the
Bit Error Rate (BER) for any given SNR, under different assumptions on
the knowledge of the CSI statistics.  We also present robust
constellation designs given imperfect knowledge of the channel
statistics.  Keeping in mind the fact that the channel moments can be
estimated on time scales larger than those needed for estimating
instantaneous phase (for which we typically need resource-consuming
training sequences), in our work we investigate how many antennas we
need for a certain BER performance. Through analysis and simulations,
we find that the suggested schemes can outperform several existing
noncoherent schemes, e.g., conventional ASK modulation, as well as
pilot-based schemes, such as PAM modulation when the coherence time of
the channel is small.

To the best of our knowledge, this line of work, with initial results
appearing in \cite{manolakos2014globecom}, is the first to consider an
energy-based encoding and decoding procedure for a noncoherent large
antenna system and demonstrate its performance gains over other
noncoherent and coherent schemes. % The framework outlined in our work
% does not make any assumption about the underlying fading
% distribution except for the fact that it should have a moment
% generating function (m.g.f.), and allows us to explicitly
% characterize the impact of the number of antennas on the BER as the
% number of antennas grows.

\subsection{Prior Work}

Noncoherent communication, i.e., communication without instantaneous
channel information, has attracted a lot of research due to either the
difficulty of channel state acquisition, or the need for low hardware
complexity and low energy consumption. The earliest incarnations of
noncoherent systems were mostly motivated by the simplicity of the
receiver circuitry: the use of envelope detectors can be traced back
to the well-studied quadrature and square law receivers
\cite{brehler2001asymptotic,kam1995generalized} employed in the
noncoherent detection of several well-known modulation schemes, such
as Frequency-Shift Keying (FSK), Amplitude Shift Keying (ASK)
\cite{kim2012analysis} and Pulse Position Modulation (PPM)
\cite{carbonelli2006m}. However, the spectral efficiency of these
systems is inferior to that of their coherent counterparts. Hence,
with the paucity of spectrum resources in cellular and with the
advances in device manufacturing, coherent systems using phase
acquisition circuitry at the receivers mostly replaced noncoherent
systems.

However, as the demand for mobile data in wireless broadband
communications increases dramatically every year, there is an evident
trend towards higher and higher carrier frequencies and large antenna
arrays \cite{rusek2013scaling}, \cite{rappaport2013millimeter}. In
these scenarios, the issues of simple circuit designs, inexpensive
hardware components and energy efficiency become as crucial to system
design as spectral efficiency
\cite{rusek2013scaling,larsson2013massive,bjornson2013massive}. While
the number of RF chains goes up with an increasing number of antennas,
thereby causing increased complexity and energy consumption, hardware
impairments such as phase noise and I/Q imbalances also become more
severe at both the transmitter and the receiver.  Architectures which
use simple, robust and energy efficient designs are thus important to
realizing many of the performance benefits of massive MIMO.  % Spatial
% Multiplexing (SM) \cite{mesleh2008spatial,di2014spatial} is one
% example of a promising system which has only one RF chain.  However,
% even there, we have several important challenges, such as fast
% antenna switching, small directional beamforming gain and the need
% for accurate CSI at the receiver.

% proposes a noncoherent communication system that uses the (GLRT) to
% jointly recover the channel and the transmitted symbols, whenever
% one wants to avoid the estimation of the long-term statistics of the
% channel.  Interestingly, even though the GLRT decoder has in general
% worse performance than the ML decoder, it is identical to the latter
% for unitary signaling and i.i.d. fading
% \cite{brehler2001asymptotic}.  A similar line of work studying the
% effect of non-instantaneous channel state information on the
% beamforming gain in multi antenna systems can be found in
% \cite{akdeniz2013millimeter} and \cite{lozano2007long}.

Initial investigations on the rate loss incurred by even a small
training overhead in a massive SIMO system show that in several
scenarios of low SNR, high mobility, or a large line of sight (LOS)
component, a noncoherent system achieves a better probability of bit
error than a coherent system for the same effective rate
\cite{mainakicc2015}.  Thus, noncoherent schemes seem to be a
competitive alternative to coherent schemes in the design of practical
large-antenna systems.

Within this line of research, \cite{mccloud2002signal} focuses on the
noncoherent ML decoder and proposes signal constellation designs using
a metric motivated by a union bound on the probability of error in the
high SNR regime. Similar metrics, also motivated by a high SNR
analysis, are presented in \cite{barg2002bounds} where the worst-case
chordal distance is employed to place the codewords as far apart as
possible.  An alternative to the ML detection of the transmitted
symbols is considered in \cite{warrier1999noncoherent}, where the
authors consider the problem of joint channel and transmitted symbol
estimation, and propose a minimum distance criterion for code design
using the Generalized Likelihood Ratio Test (GLRT) in the AWGN channel
at high SNR.  The GLRT is necessary when the channel statistics are
not known precisely and it involves maximizing the noncoherent
likelihood function over all possible values of the channel
statistics.  While these decoders are applicable to a very general
class of channels, in this work, our focus is on understanding and
optimizing the performance of energy based decoders with an increasing
number of antennas.

The rest of the paper is organized as follows. We present the system
model in Section \ref{sec:model}, and summarize relevant work on its
asymptotic characterization in Section \ref{sec:constellation}.  Then,
Section \ref{subsec:overview} presents the constellation design
problem and Section \ref{subsec:Perfect} describes the solution to
this problem when the channel distribution is perfectly known.
Section \ref{subsec:4moments} shows how the constellation design
problem can be done when only the first few moments of the fading
distribution are known, and Section \ref{subsec:robust} addresses the
case when even the moments are not known perfectly. Finally, in
Section \ref{sec:numerical}, we present plots showing the performance
of the suggested schemes with representative statistics. Section
\ref{sec:conc} summarizes this work.

\subsection{Notation}
We use $[k]$ to denote the set $\{1,2,\cdots,k\}$ where $k$ is an
integer. $\mathbb{C} ^{n \times m}$ is the set of all complex-valued
matrices of size $n \times m$.  For a matrix $\Hmat \in \mathbb{C} ^{n
  \times m}$, the $(i,j)$-th element is denoted by $H_{i,j}$ and for a
vector $\hvec \in \mathbb{C} ^{n \times 1}$, the $i$-th element is
denoted as $h_i$. $\re(\cdot)$ and $\im(\cdot)$ represent the real and
imaginary terms, respectively.  $\mathcal{CN}(\mu,\Rmat)$ represents
the distribution of circularly symmetric complex Gaussian (CSCG)
random vectors with mean vector $\mu$ and a covariance matrix $\Rmat$.
The symbol $\triangleq$ is used to denote a definition.
$\symbolconstellation$ refers to a set of power levels that the
transmitter uses and $n$ is used to denote the number of receive
antennas.  We refer to a $(K,\gamma)$ Rician fading channel as a
channel with Rician fading ($K$-factor in dB units and unit second
moment) and additive Gaussian noise with power $\sigma^2=-\gamma$ in
dB.

% and $\doteq$ to denote equality to the first order in the exponent,
% i.e., $a_n \doteq b_n$ means that $\lim_{n\rightarrow
% \infty}\frac{1}{n} \log\lb \frac{a_n}{b_n} \rb=0.$ We use $U$ to
% refer to a random variable, and $u$ to a realization of the same.
% $\expectation \lsb U\rsb$ denotes the expectation of a random
% variable $U$.

\section{System Model}
\label{sec:model}
Consider one single antenna transmitter in a flat fading channel and a
receiver with $n$ antennas, where $n$ is a large (but finite) number.
The system is represented as
\begin{align}
  \yvec = \hvec x + \vvec,
\end{align}
with $\yvec \in \mathbb{C}^{n\times1}$, $x \in \mathbb{C}$, $\vvec \in
\mathbb{C}^{n\times1}$, $\hvec \in \mathbb{C}^{n \times 1}$ and each
$v_i \sim \mathcal{CN}(0,\sigma^2)$, $h_i \sim f(h)$, such that
$\expectation [h_{i}]=\mu, ~\expectation \lsb |h_i - \mu|^2
\rsb=\sigma_h^2$, and $f(h)$ is the probability density function of
the channel distribution. For normalization purposes and for
notational simplicity, we also assume that $\expectation[|h_i|^2]=1$
and $\expectation[|x|^2]=1$ so that parameters such as long-term
shadowing, path-loss and antenna gain are incorporated in the
$\sigma^2$. Then, the average SNR per antenna at the receiver for this
model is $$\gamma \triangleq \frac{\expectation[|h_i|^2]}{\sigma^2} =
\frac{1}{\sigma^2}.$$ We further assume that the density function
$f(h)$ is such that, for any fixed $x \in \mathbb{C}$, the moment
generating function of $|y_i|^2$, i.e.,
$\expectation[e^{\theta|y_i|^2}]$, exists and is twice differentiable
in an interval around $\theta = 0$.  Many fading distributions fall
within this model, e.g., Rayleigh and Rician fading
\cite{goldsmith2005wireless}, in which case $h_{i} \sim
\mathcal{CN}(\mu,\sigma_h^2)$.

Note that an important aspect of this system model, similar to many
works in the massive MIMO literature, is the assumption that the
channel realizations across the antennas are i.i.d. random variables.
While this assumption is not typically accurate in practice, some
recent measurements
\cite{hoydis2012channel,gao2012measured,gao2014massive} suggest that,
despite the statistical difference between the measured multiple
antenna channels and the i.i.d. channels, most of the theoretical
conclusions made under the independence assumption are still valid in
real massive MIMO channels.

Motivated by our recent asymptotic results in \cite{mainak2014,
  mainak2014journal, mainakITW2015}, this work focuses on
symbol-by-symbol encoding and decoding schemes that use an average
energy-based transmitter and receiver design. This means that
information is modulated in the power of the transmitted symbols,
$|x|^2$, and the receiver estimates only the average power of the
received signal, $\frac{||y||^2}{n}$. % We are interested in those
% scenarios where coherent communication is not possible, e.g.,
% scenarios in which the channel varies so fast due to small-scale
% fading that it is probably only possible to estimate well the first
% few moments of the channel distribution but not the exact channel
% realization at the receiver. We first describe the details of the
% average energy-based encoder and decoder, and then we motivate the
% usage of this specific architecture.
We describe this next.

\subsection{Transmitter architecture: energy encoder} 
The transmitter encodes information only in the \textit{power} of the
transmitted symbols, i.e., it transmits symbols with power levels from
a codebook $$\symbolconstellation = \{p_{1},p_{2},\cdots,p_{L}\},$$
where $p_k \in \mathbb{R_+},$ subject to an average power constraint
$$\frac{1}{L}\sum_{k=1}^{L}p_{k}\leq 1,$$ assuming equiprobable
signaling. Here $p_{k}\in \mathcal{P}$ is the power level of the
$k^{th}$ symbol and $L$ is the cardinality of $\symbolconstellation$.
Note here that we do not encode information in the
phase.% Obviously, any set of transmitted symbols with powers
% that belong in codebook $\symbolconstellation$ are equivalent. Also,
% note that in this work constellation point refers to the power of
% the corresponding transmitted symbol. Contrary to typical modulation
% techniques, which usually specify the amplitude and the phase of the
% transmitted symbols, we only describe how the powers of the
% transmitted symbols should be chosen.

\subsection{Receiver architecture: energy decoder}
Assume the user transmits a symbol whose power is the $k^{th}$
constellation point from $\symbolconstellation$, i.e, $p_k$. In order
for the receiver to detect $p_k$, it only computes the following
statistic
\begin{align}
  \label{eq:stat}
  \frac{\|\mathbf{y}\|^2}{n} = \frac{\sum_{i=1}^n|y_i|^2}{n} \in
  \mathbb{R}^+,
\end{align} i.e., it estimates only the average received power across
all its antennas. Based on its knowledge of the statistics of the
channel, the receiver divides the positive real line into
non-intersecting intervals or \emph{decoding regions} $$\mathcal{D} =
\{\mathcal{I}_{k}\}_{k=1}^L,$$ (each $\mathcal{I}_k$ corresponding to
each ${p_k} \in \symbolconstellation$), and returns
\begin{align}
  \label{eq:energydecoder}
  \hat {k} \in \left\{ \tilde{k}: \frac{\|\mathbf{y}\|^2}{n} \in
    \mathcal{I}_{\tilde k} \right\}.
\end{align}
% Then, we refer to the {\color{blue} Can we refer to the
% constellation and decoding regions separately to prevent confusion
% ?}  \textit{constellation} $\mathcal{C}$ as the set that contains
% the codebook $\mathcal{P}$ and the corresponding decoding regions
% $\{\mathcal{I}_k\}$, i.e., $\mathcal{C} =
% \{\mathcal{P},\mathcal{I}_1,\cdots,\mathcal{I}_{L}\}.$ The
% constellation $\mathcal{C}$ is decided by the system prior to the
% start of the communication based on the statistics on the channel.

The probability of error $P_e(p_k)$ when the $k^{th}$ power level $p_k
\in \mathcal{P}$ is transmitted, and the average Symbol Error Rate
(SER) $P_s$ for any fixed constellation size $L$ are defined as
\begin{align}
  \label{eq:pe}
  P_e (p_k) \triangleq Pr\{\hat{k} \neq k \}, ~~P_s \triangleq
  \frac{1}{L}\sum_{k = 1}^L P_e(p_k),
\end{align}
respectively, assuming equiprobable signaling.

\subsection{Discussion}
The use of energy-detection-based transmission and decoding is
motivated by the fact, proved in \cite{mainak2014journal}, that such
an encoding and decoding method achieves the same BER as a noncoherent
maximum likelihood (ML) scheme in the Rayleigh fading channel. To see
this, as explained in more details in \cite{mainak2014journal}, assume
the transmitter sends a symbol $x\in\mathbb{C}$. The log likelihood
function for a noncoherent Rician fading channel, i.e., $h_{i} \sim
\mathcal{CN}(\mu,\sigma_h^2)$, is $$\log f_{x}^{NC} (\yvec) =
\frac{\|\yvec- \mu x\ones\|^2}{\sigma + \sigma_h^2 |x|^2} + n \log
\lb \pi (\sigma + \sigma_h^2|x|^2)\rb,$$ and therefore, the
\textit{noncoherent ML decoder} is
\begin{align}
  \label{eq:noncoh}
  \hat k = \argmax_{x: |x|^2 =p_k, \forall k} \log f_{x}^{NC} (\yvec).
\end{align}
For $\mu=0$, i.e., Rayleigh fading, the noncoherent ML decoder depends
only on $\|\yvec\|^2$. Thus for suitably chosen decoding regions
$\mathcal{D}$, it performs as well as the ML decoder.  In general for
$\mu \neq 0$, energy based detectors are not optimal.  However, as
shown in Figure \ref{FIgA} for representative values for $\mu$,
the SER performance gap is typically small.

The proposed architecture requires a very simple one-dimensional
statistic of the received signals, which allows for a simplified RF
chain design.  A general noncoherent ML or coherent detector, on the
other hand, requires much more complicated circuits.

% In the sequel, we first summarize the analytical upper bound on SER
% proved in \cite{mainak2014journal}, which is then used to present
% the suggested constellation design algorithms.

\section{Constellation Design Optimization}
\label{sec:constellation}
We are interested in the problem of minimizing SER for any fixed
constellation size $L$ and fixed $n$, i.e.,
\begin{equation}
  \label{eq:constprobl}
  \begin{aligned}
    \mathcal{P}^*,\mathcal{D}^* =  & {\text{ argmin}} ~~~~~~~ \frac{1}{n}\log(P_s)  \\
    & \mbox{   subject to}~~~~ \frac{1}{L}\sum_{k=1}^{L}p_k \leq 1, \\
    & ~~~~~~~~~~~~~~~~~ 0 \leq p_k, ~\forall k \in [n]. \\
  \end{aligned}
\end{equation}
The optimization above is over all codebooks $\mathcal{P}$ and
decoding regions $\mathcal{D}$.  Unfortunately, this is in general a
difficult problem to solve. The scope of this work is to solve a
specific relaxation of this problem motivated by the large $n$
asymptotics ($n$ is large but finite) for any channel distribution
that has a m.g.f. Specifically, we consider maximizing the error
exponent of SER with respect to $n$, or a second-order approximation
of it.  Before describing the relaxation we first define the notion of
the error exponent of SER with respect to $n$. Using this asymptotic
characterization not only helps us make the constellation design
problem tractable, but also gives us designs which work for little
prior statistical characterization and decouples the effect of the
number of receive antennas from the effect of the channel
distribution.

\subsection{Error exponents}
\label{sec:upper}

Fix any codebook $\cal P$. Define the \emph{receiver's constellation
  points} $r(p_k)$ to be the value of the average received energy when
the transmitter sends the $k^{th}$ power level, i.e., $$ r(p_k)
\triangleq p_k +\sigma^2.$$ We now observe that
\begin{align*}
  \frac{||\mathbf{y}||^2}{n} = & \frac{||\mathbf{h} \sqrt{p_k}
    +\vvec||^2}{n} = \frac{||\mathbf{h}||^2}{n} p_k +
  \frac{||\vvec||^2}{n} + 2\frac{
    \textrm{Re}(\mathbf{h}^*\vvec)}{n}\sqrt{p_k},
\end{align*}
so, in the limit of large $n$, due to the law of large numbers and the
independence of $\hvec$ and $\vvec$, it follows
that $$\lim\limits_{n\rightarrow \infty}\frac{||\mathbf{y}||^2}{n} =
r(p_k).$$ For finite $n$, the statistic $ \frac{\|\mathbf{y}\|^2}{n}$
would deviate from the value $r(p_k)$. To characterize this deviation,
it is helpful to define
\begin{equation}
  \label{eq:U}
  u_{k,i} = \left | h_{i} \sqrt{ p_{k} } + \nu_i\right|^2 - E\left[\left | h_{i} \sqrt{ p_{k} } + \nu_i\right|^2 \right ] = \left | h_{i} \sqrt{ p_{k} } + \nu_i\right|^2 - r(p_k)
\end{equation}
as the random variation of the received energy at the $i^{th}$ antenna
around its expected value $r(p_k)$. Note that $\{u_{k,i}\}_{i=1}^n$
are independent realizations of the same zero-mean random variable
$U_k \sim g_k(u)$ whose m.g.f. is
\begin{align}
  \label{eq:Mk}
  M_k(\theta) \triangleq \expectation[e^{\theta U_k}].
\end{align}
In the above, $M_k(\theta)$ depends on the statistics of the channel
and the noise, and the power level $p_k$.

In \cite{mainak2014journal}, we derived an upper bound for the
objective in \eqref{eq:constprobl} as follows:
\begin{align}
  \label{upper}
  P_s\leq \frac{1}{L} \sum_{k=1}^L \lb e^{-nI_{R,k}(d_{R,k})}
  +e^{-nI_{L,k}(d_{L,k})} \rb,
\end{align}
where
\begin{align}
  I_{L,k} (d) \triangleq \sup_{\theta\geq 0} \left(\theta d -
    \log(M_k(-\theta))\right), ~~~I_{R,k} (d) \triangleq
  \sup_{\theta\geq 0} \left(\theta d - \log(M_k(\theta))\right),
\end{align} are defined as the \textit{left} and \textit{right rate
  functions} of $p_k$ respectively. In addition, $\sup_{x\in A}f(x)$,
is the supremum of
$f(x)$ in $A$.  The decoding regions are chosen as $${\cal I}_{k}
\triangleq %In other words, for every constellation point $p_k$, the values $d_{R,k}$ and $d_{L,k}$ define .
\left(r(p_k)-d_{L,k},r(p_k)+d_{R,k}\right],$$ for some $d_{L,k}$ and
$d_{R,k}$.  Define \[I_{k} \triangleq \min \big(I_{L,k} (d_{L,k}),
I_{R,k}(d_{R,k})\big ) \] to be the \textit{rate function} of the
constellation point $p_k$. Then, it was shown in
\cite{mainak2014journal} that
\begin{align}
  \label{eq:worst_error}
  I_e \triangleq \lim_{n \rightarrow \infty} -\frac{\log(P_s )}{n} =
  \min_{k \in [L]}
  I_{k}, %\mbox{ and that } P_e \leq P_U \triangleq 2e^{-n I_e}
\end{align}
i.e., the \textit{error exponent of SER}, denoted as $I_e$, is the
same as the worst rate function of the constellation points. In other
words, for $n$ large enough, the probability of error performance is
dominated by the constellation point with the worst rate
function. Therefore, the constellation points $\mathcal{P}$ and the
corresponding decoding regions $\mathcal{D}$ could be chosen in such a
way as to maximize the error exponent of SER, i.e., (error exponent
maximization problem)
\begin{equation}
  \label{eq:constprobl2}
  \begin{aligned}
    & {\text{maximize}}  ~~~~~~~I_e  \\
    & \mbox{subject to} ~~~~~~\frac{1}{L}\sum_{k=1}^{L}p_k \leq 1,~ 0 \leq p_k. \\
  \end{aligned}
\end{equation}
Observe that problem \eqref{eq:constprobl2} is a relaxation of problem
\eqref{eq:constprobl} for any finite $n$, since its objective function
is an upper bound on the objective function of the latter. Yet, note
that due to \eqref{eq:worst_error} this relaxation would provide an
asymptotically optimal constellation with increasing $n$, even if it
does not explicitly solve the SER minimization problem exactly for any
finite $n$ \eqref{eq:constprobl}. Furthermore, an interesting aspect
of \eqref{eq:constprobl2} is that it depends only on the channel
distribution, and not on the number of antennas.  By decoupling number
of antennas from the channel distribution it allows us to study
achievable SERs with large $n$ without explicitly solving
\eqref{eq:constprobl}.

In \cite{mainak2014journal} we showed the following about the left and
right rate functions for any $p_k$:
\begin{lemma}
  \label{lemma2}
  The left and right rate functions $I_{R,k}(d),I_{L,k}(d)$,
  respectively, of the power level $p_k$ enjoy the following
  properties:
  \begin{itemize}
  \item They satisfy \[\lim_{d\rightarrow 0} \frac{I_{R,k}(d)}{d^2}
    =\lim_{d\rightarrow 0}
    \frac{I_{L,k}(d)}{d^2}=\frac{1}{2\expectation [U_k^2]},~~\mbox{
      where }~~ U_k = |h\sqrt{p_k}+v|^2-p_k-\sigma^2,\] with $h \sim
    f(h)$ and $v \sim \mathcal{CN}(0,\sigma^2)$.
  \item They are non-negative, convex and monotonically increasing for
    positive $d$ for a fixed non-negative $p_k$, and monotonically
    decreasing for non-negative $p_k$ for a fixed positive $d$.
  \item It holds that $I_{L,k}(0) = I_{R,k}(0) = 0$ for any
    non-negative $p_k$.
  \end{itemize}
\end{lemma}

The above lemma provides important insights into the dependence of the
rate functions on the power levels and channel
statistics. Specifically, for a small $d$, which practically means
large constellations, we can approximate the rate function using only
the first few moments of the channel distribution ($E[U_k^2]$ depends
on the first, second, and forth moment of the fading distribution as
we explicitly show in Section \ref{subsec:4moments}). Also, for a
fixed $d$, increasing $p_k$ leads to smaller rate functions, i.e.,
worse SER performance, or in other words, the constellation points
that correspond to high power levels generally experience worse
performance than those with low power levels for decoding regions of
the same size.  We now describe how these analytical properties could
be used to solve for the maximum error exponent as defined in problem
\eqref{eq:constprobl2}.
% Actually, this exact behavior of the rate functions is exploited in
% our constellation designs: space the power levels onto the positive
% real line in such a way such that all the constellation points
% experience the same rate function. Then, we can guarantee that the
% proposed design has a positive error exponent with a large but
% finite $n$, and explicitly characterize the dependence of the
% achieved probability of error performance as a function of $n$.

% For a Rician fading channel it holds that $ \expectation[U_k^2] =
% \sigma^4 + 2 p_k \sigma^2 + \frac{(1+2K)p_k^2}{(1+K)^2}$.

\section{Constellation designs}
\label{sec:single_perfect_exact}

\subsection{Overview}
\label{subsec:overview}

In this section we consider three cases of constellation designs to
maximize the error exponent in \eqref{eq:constprobl2}, where each case
corresponds to a different assumption on the availability of
statistical information about the
channel.% We start from problem \eqref{eq:constprobl2}.
% we now con discuss in detail three different scenarios of
% information available to the encoder and the decoder which lead to
% different constellation design problems:
\begin{itemize}
\item Case 1: Subsection \ref{subsec:Perfect} presents a design which
  assumes that the encoder and decoder know perfectly the channel
  distribution. This constellation is denoted as
  $\symbolconstellation_{K,\gamma}$.
\item Case 2: Subsection \ref{subsec:4moments} presents a design in
  which only the first four moments of the channel distribution, are
  perfectly known. We will show later that these four moments suffice
  for a near-optimal design at high rates. This constellation is
  denoted as $\hat\symbolconstellation_{K,\gamma}$.
\item Case 3: Subsection \ref{subsec:robust} presents a design in
  which even the latter are imperfectly known. The corresponding
  constellation is $\hat\symbolconstellation^{(a)}_{K,\gamma}$, where
  $a$ is the uncertainty in dB around the nominal values $K$ and SNR.
\end{itemize}
In addition to the constellation designs above, we also consider a
minimum distance constellation design, denoted as
$\symbolconstellation^{min}$, that was proposed in
\cite{mainak2014journal}, which has $$p_k = \frac{2(k-1)}{L-1}, \mbox{
  for } k \in [L]$$ with decoding regions $$\mathcal{I}_0 =
\left[0,\frac{1}{L-1}\right],~ \mathcal{I}_k =
\left[\frac{2k-3}{L-1},\frac{2k-1}{L-1}\right]~\forall k \in
\{2,\cdots,L-1\}, ~\mathcal{I}_L=
\left[2-\frac{1}{L-1},\infty\right].$$ We now describe each of the
above constructions.
% We denote as $\symbolconstellation^{min}$ a minimum distance
% constellation design that was proposed in
% \cite{mainak2014,mainak2014journal}. This is close to optimal only
% under a high noise assumption. The new approach presented in this
% work generalizes the minimum distance design criterion to very
% general scenarios without constraints on the SNR. Furthermore, as a
% byproduct of the above designs, it is possible to propose a
% constellation in which the family of the channel distribution is
% known, but the distribution's parameters are imperfectly known.

% Since this design follows the same principles as presented in the
% subsequent sections, a description of this design is not provided
% for the sake of
% brevity. %Numerical results for this design appear in Section \
% ref{sec:numerical}.  The suggested constellation design problem
% provides asymptotically optimal designs with respect to the error
% exponent for any channel distribution whose m.g.f. exists and is
% twice differentiable in an interval around zero.

\subsection{Perfect knowledge of channel distribution}
\label{subsec:Perfect}

We first discuss the constellation design with perfect knowledge of
the channel distribution at the receiver.  Since the exact channel
distribution is known, $M_k(\theta)$ is also known at the receiver and
transmitter for any chosen $p_k$. Then, \eqref{eq:constprobl2} can be
written as
\begin{equation}
  \begin{aligned}
    \label{eq:relaxedproblem}
    & \underset{ \{p_k , d_{L,k} , d_{R,k}\}_{{k} \in [L] }}{\text{maximize}} & & \min_{k \in [L]} \big( I_{L,k} (d_{L,k}),  I_{R,k}(d_{R,k})\big ) \\
    &&& ~0\leq p_{1}< p_2 < \cdots < p_{L}, \\ &&& d_{L,k}\geq 0, ~d_{R,k} \geq 0,\\
    &&& \frac{1}{L}\sum_{k=1 }^{L}p_{k} \leq 1.
  \end{aligned}
\end{equation}
assuming decoding regions of the form ${\cal I}_k =
(r(p_k)-d_{L,k}~,r(p_k)+d_{R,k}],$ where for simplicity we assume that
$d_{L,1}= d_{R,L}=\infty$.

\begin{figure}[ht]
  \centering
  \includegraphics[width=10cm]{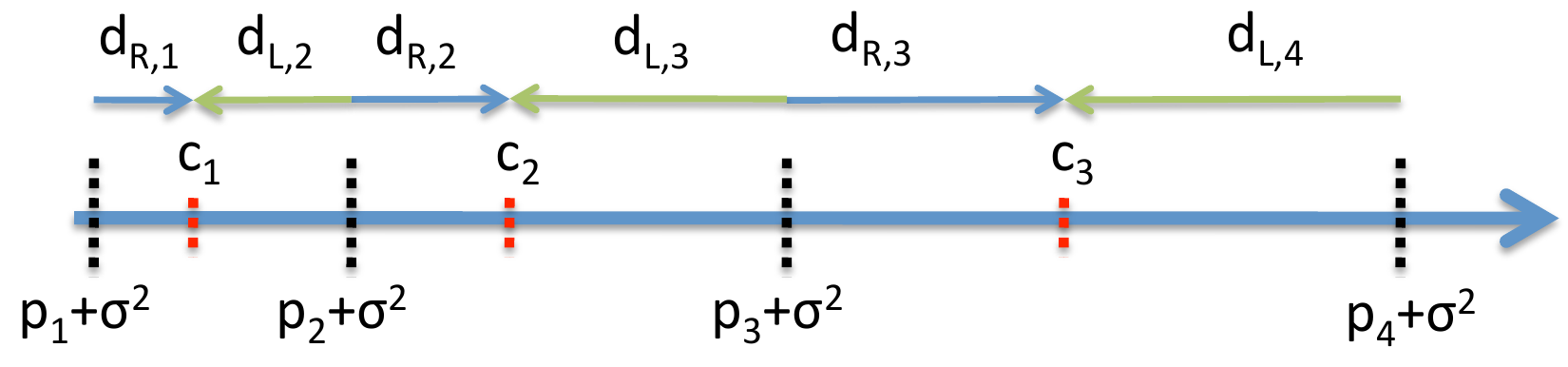}
  \caption{Example of the decoding regions using $L=4$}
  \label{fig:figure1}
\end{figure}

\begin{algorithm}[t!]
  \small
  \caption{Bisection algorithm}
  \begin{algorithmic} \STATE {\bf function}: $[t^*,
    \symbolconstellation^*]$ = {\bf Bisection}( ) \STATE $t_l = 0, t_u
    = \infty$ \REPEAT \STATE $t = \frac{t_l+t_u}{2}$ \STATE
    $[\symbolconstellation, S_t]$ = \textbf{ConstellationDesign}( t )
    \STATE \textbf{If} ~~$(S_t < 1)$: $t_l = t$;~~ \textbf{else}
    ~~$t_u = t$ \UNTIL{$|t_u-t_l|<\epsilon$ and $|S_t - 1|<\epsilon$}
    \RETURN $\symbolconstellation^* = \symbolconstellation;$ $t^* =
    t;$
  \end{algorithmic}
\end{algorithm}
Algorithm \ref{algo1} describes in detail how to get the solution of
the optimization problem \eqref{eq:relaxedproblem} and a detailed
proof is presented in Appendix \ref{appedA}. To exemplify the
procedure and provide an intuitive argument for the validity of the
suggested construction we consider the case with $L=4$ shown in
Figure \ref{fig:figure1}.  The design is based on the following two
properties that follow from Lemma \ref{lemma2}:
\begin{enumerate}
\item Both $I_{L,k}(d)$ and $I_{R,k}(d)$ are non-negative and
  monotonically increasing functions of $d$ for a fixed $p_k$. This
  means that increasing the size of the decoding regions always helps
  to increase the resulting rate functions, and therefore increase the
  minimum amongst them.
\item Both $I_{L,k}(d)$ and $I_{R,k}(d)$ are monotonically decreasing
  functions of $p_k$ for a fixed $d$.
  % This means that transmitting with low power levels should be
  % always preferred.
\end{enumerate}

Based on these two properties, we have the following sequential
construction: assume there exists a constellation with error exponent
$t^*$ that satisfies the power constraint. This means that the left
and right rate functions of all the constellation points at the
receiver are at least $t^*$. To find this constellation choose first
the minimum possible value for $p_1$. Then, choose the boundary of the
decoding region to the right of $r(p_1)=p_1+\sigma^2$, i.e., $c_1$, as
show in Figure \ref{fig:figure1}, such that the right rate function of
$r(p_1)$, $I_{R,1}(c_1-r(p_1))$, is at least $t^*$ on the
boundary. Then, choose the smallest $p_2$ such that $r(p_2)>c_1$ and
the left right rate function of $r(p_2)$, $I_{L,2}(r(p_2)-c_1)$, is at
least $t^*$. Note that choosing a higher $p_2$ is always an option but
this will lead to a design that uses more power than necessary. We
perform this procedure sequentially until we find $p_L$. Then we check
if the average power constraint is satisfied. If that is the case, the
assumption that there exists a constellation with error exponent at
least $t^*$ that satisfies the power constraint was correct. If not,
we should discard this constellation, decrease $t^*$ and repeat the
procedure. Proof that this procedure gives the optimal error exponent
is presented in Appendix \ref{appedA}.

\begin{figure}
  \begin{minipage}{1\textwidth}
    \centering \subfigure{
      \epsfig{figure=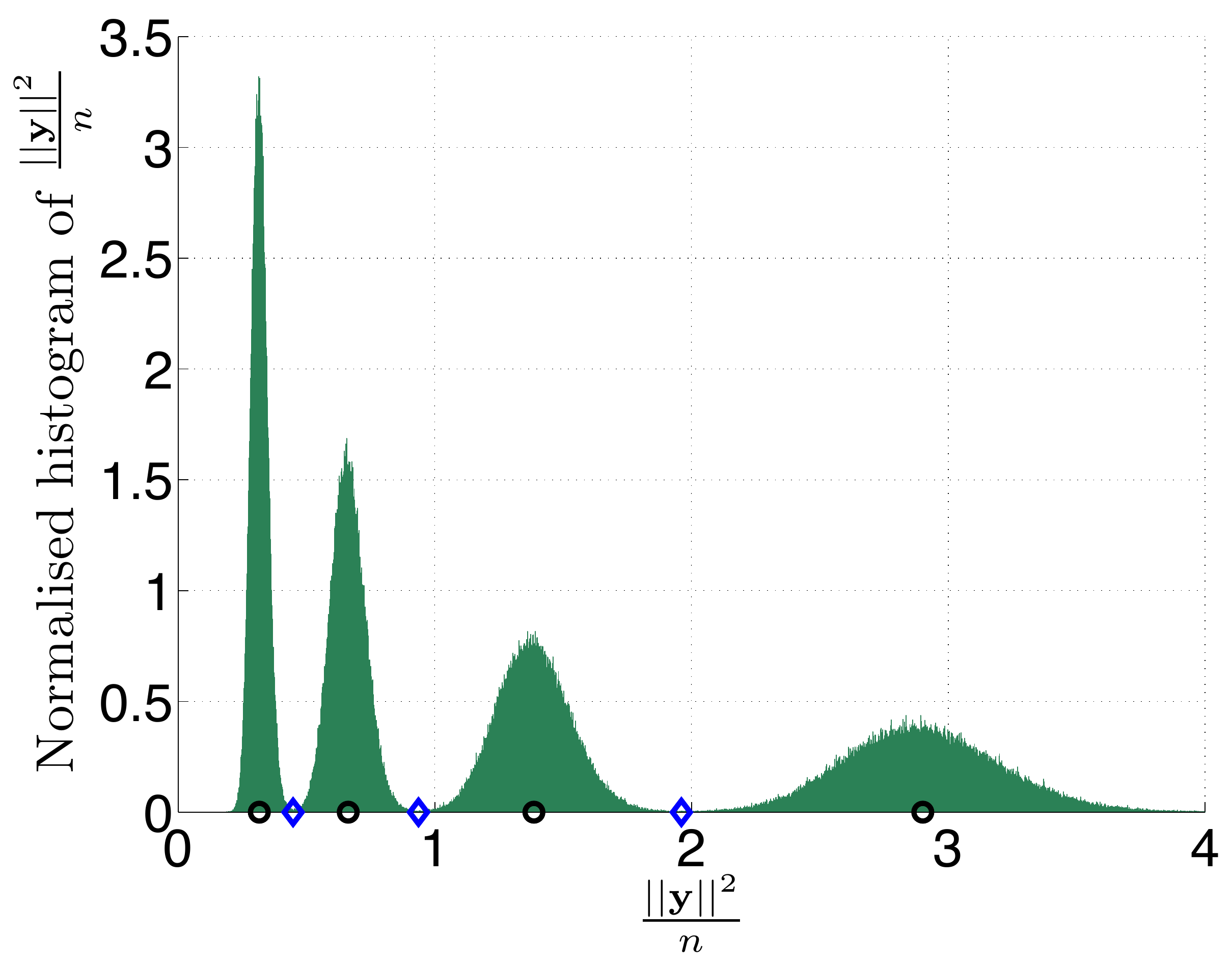,width=7cm}}
    \subfigure{
      \epsfig{figure=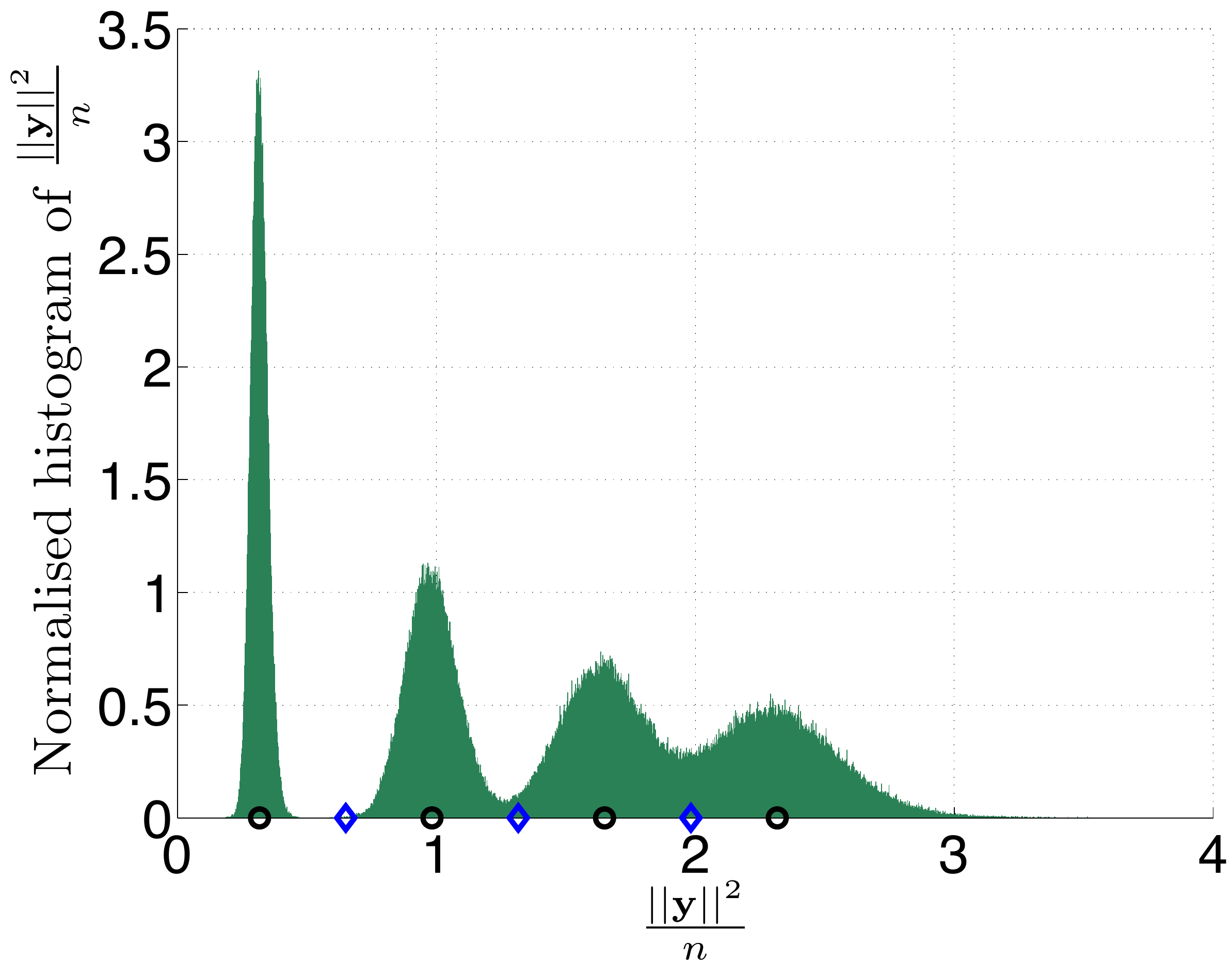,width=7cm}}
    \caption{Example of the histogram of $\frac{||\yvec||^2}{n}$ for
      $n=100$ antennas for $\symbolconstellation_{K,\gamma}$ and
      $\symbolconstellation^{min}$}
    \label{FigExam}
  \end{minipage}
\end{figure}

For $L=4$, Figures \ref{FigExam}-(a) and \ref{FigExam}-(b) show the
normalized empirical histogram of the received statistic
$\frac{||\yvec||^2}{n}$ in a Rayleigh fading channel of the suggested
constellation: $\symbolconstellation_{K,\gamma}$ and
$\symbolconstellation^{min}$. The circles and diamonds on the x-axis
show the $r(p_k)$ and the $c_k$ (boundaries of the decoding regions)
respectively. Observe that for the $\symbolconstellation^{min}$
constellation there is a significant overlap in the histogram and
therefore the receiver experiences significant symbol error rates.
Also, observe that as $p_k$ increases, the variation around $r(p_k)$
also increases, due to the special nature of the energy detector at
the receiver.

\begin{comment}
  In short, in order to find the solution, we need to identify the
  largest value $t^*$ for which the left and right rate functions of
  the constellation points are equal while satisfying the average
  power constraint. To do so, for any fixed $t$, we sequentially
  decide the value of $p_1$, $d_{R,1}$, $p_{2}$, $d_{R,2}$, $\cdots$,
  $p_L$ in this specific order, by ensuring that we use the minimum
  possible power levels and a minimum value $t$ for all the rate
  functions. To identify the maximum $t^*$, it is possible to perform
  a binary search since the average power is a monotonically
  increasing function of $t$.
\end{comment}

\begin{figure}
  \begin{minipage}{1\textwidth}
    \centering \subfigure[Comparison of Energy and ML decoder in
    Rician fading with $\gamma=10$ dB]{
      \epsfig{figure=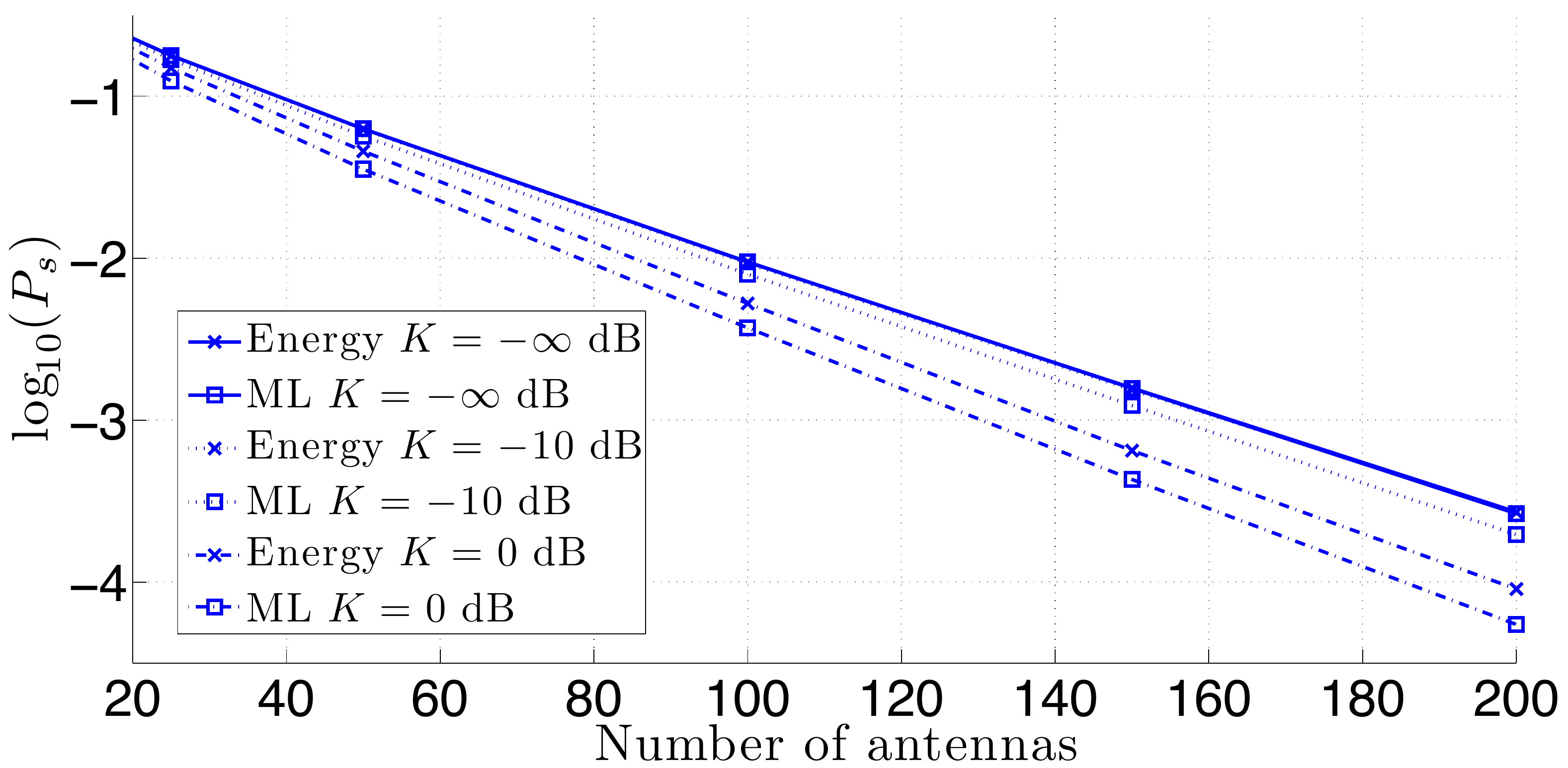,width=8.5cm}} \subfigure[ $K
    = 0$ dB, $\gamma = 5$ dB]{
      \epsfig{figure=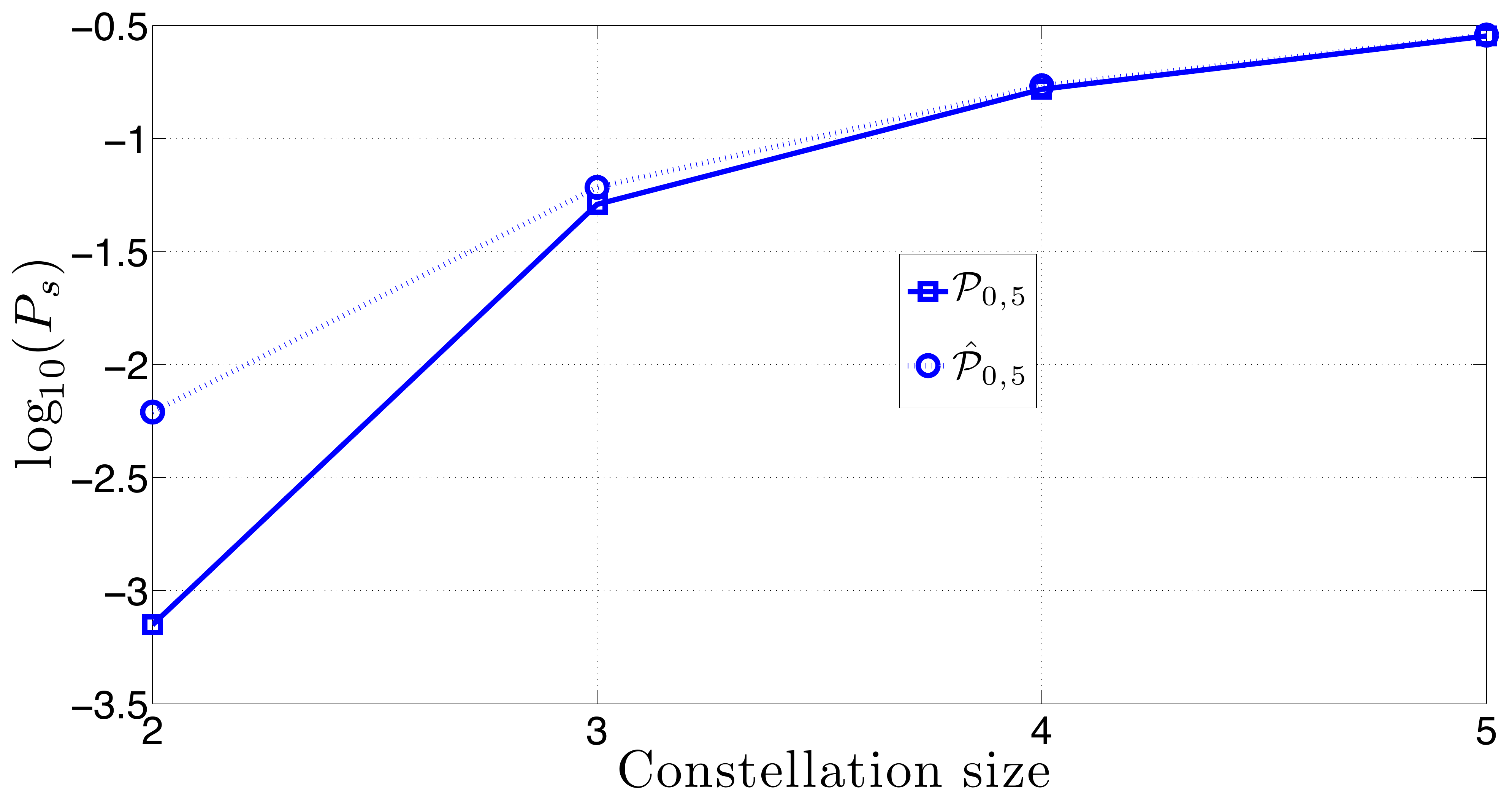,width=8.5cm}}
    \caption{(a) Comparison of the energy decoder with the noncoherent
      ML decoder in Rician fading, (b) Comparison of
      $\symbolconstellation_{K,\gamma}$,
      $\hat\symbolconstellation_{K,\gamma}$ as a function of the
      constellation size $L$ in Rician fading}
    \label{FIgA}
  \end{minipage}
\end{figure}

Furthermore, Figure \ref{FIgA} shows a numerical comparison of the SER
of two systems that transmit using the same codebook $\mathcal{P}$ in
Rician fading, but with different decoders; the first system uses the
decoder in \eqref{eq:energydecoder} and the decoding regions resulting
from the above procedure, and the second system uses the ML noncoherent
decoder \eqref{eq:noncoh}. Observe that there is no difference in the
performance of the two systems in Rayleigh fading. Moreover, even in
Rician fading with a relatively strong LOS component, the difference
in performance is small. This suggests that energy decoders can
roughly match the performance of optimal decoder structures even when the fading distribution has a non-trivial LOS component.

% \epsfig{figure=EnergyvsEnergK0.pdf,width=7.3cm}} \subfigure[Rician
% fading with $K = \{-5,0,5\}$ dB and $\gamma=5$ dB]{

\begin{algorithm}[t!]
  \small
  \caption{Constellation design: Perfect channel distribution
    knowledge} \label{algo1}
  \begin{algorithmic} \label{algo2} \STATE $~$ \STATE $[t^*,
    \symbolconstellation^*]$ = {\bf Bisection}( ); \STATE
    $~$\line(1,0){450}
  \end{algorithmic}
  \begin{algorithmic}
    \STATE {\bf function}: $\symbolconstellation$ = {\bf
      ConstellationDesign}( t ) \STATE $p_1 = 0;$ $d_{L,1} = \infty;$
    $d_{R,L} = \infty;$ \FOR{$k=1, 2, 3, \cdots, L$} \IF{$k=1$} \STATE
    $p_1=0$ \ELSE \STATE Find the smallest $p_k>p_{k-1}+d_{R,k-1}$
    such that $ I_{L,k}(p_{k}-p_{k-1} - d_{R,k-1}) = t$ \STATE
    $d_{L,k} = p_k-p_{k-1}-d_{R,k-1}$
    \ENDIF
    \IF{$k \neq L$} \STATE Find the smallest $d_{R,k}>0$ such that
    $I_{R,k}(d_{R,k}) = t$
    \ENDIF
    \STATE $\mathcal{I}_k =
    [p_k+\sigma^2-d_{L,k},p_k+\sigma^2+d_{R,k}]$
    \ENDFOR
    \STATE $\mathcal{P} = [p_1,p_2,\cdots,p_L], S_t =
    \frac{1}{L}\sum_{k=1}^L p_k$ , $\mathcal{D} = \{
    \mathcal{I}_k\}^L_{k=1}$\RETURN $\symbolconstellation,
    \mathcal{D}, S_t$ \STATE {\bf end} {\bf ConstellationDesign}
  \end{algorithmic}
\end{algorithm}

\subsection{Perfect knowledge up to the forth channel moment}
\label{subsec:4moments}

The constellation design presented above assumes that the receiver
knows exactly the channel statistics. This may not be realistic in a
practical scenario. In this section, we relax this assumption and
consider the scenario in which the encoder and decoders only know the
first few moments of the channel distribution, i.e., up to the fourth
channel moment. The latter is especially important since determining
the exact small-scale fading channel models in some cases, such as
millimeter wave frequencies, is still ongoing research, and may not be
reliably known beyond the first few moments. Moreover, the first few
moments of the underlying channel distribution could potentially be
estimated more efficiently (i.e., on large time scales) without the
need for resource-consuming training sequences.

To see why the knowledge of the first four moments is enough to design good constellations, let $h = h_{re}+jh_{im}$, where $h_{re},h_{im} \in
\mathbb{R}$. Using Lemma \ref{lemma2} leads to
\begin{align}
  \label{eq:approxI}
  I_{R,k}(d_{R,k}) \approx \tilde I_{R,k}(d_{R,k}) \triangleq
  \frac{d_{R,k}^2}{2s(p_k)}, ~~I_{L,k}(d_{L,k}) \approx \tilde
  I_{L,k}(d_{L,k}) \triangleq \frac{d_{L,k}^2}{2s(p_k)},
\end{align}
for small $d_{R,k}$ and $d_{L,k}$, with $s(p_k) \triangleq
\expectation [U_k^2] = \alpha_1p_k^2+\alpha_2p_k+ \alpha_3$, and
\begin{align}
  \label{eq:a1}
  \negmedspace \alpha_1 \triangleq \expectation
  [h_{re}^4]+\expectation
  [h_{im}^4]+2\expectation[h_{re}^2]\expectation[h_{im}^2]-1,
  \nonumber~\alpha_2 \triangleq 2\sigma^2, ~\alpha_3 \triangleq
  \sigma^4.
\end{align}
These expressions follow from the Gaussianity of the noise and the
fact that the noise and the channel are independent and that the noise
is zero mean.  Observe that this approximation depends only on the
first, second and fourth moment of the channel distribution. For
example, in the case of Rician fading with K-factor equal to $K$ and
second moment equal to $1$, it can be shown that
$$\expectation[U_k^2] = \sigma^4 + 2 p_k \sigma^2 +
\frac{(1+2K)}{(1+K)^2}p_k^2,$$ which means that $\alpha_1 =
\frac{1+2K}{(1+K)^2}.$
% That is, in Rician fading, both sides need to know just the power of
% the LOS component and SNR, and still use the approach that we
% describe in this section.

Substituting the objective function of \eqref{eq:relaxedproblem} with
\eqref{eq:approxI} leads to the following optimization problem
\begin{equation}
  \begin{aligned}
    \label{eq:relaxedproblem3}
    & \underset{ \{p_k , d_{L,k} , d_{R,k}\}_{{k} \in [L] }}{\text{maximize}} & & \min_{k \in [L]} \big( \tilde I_{L,k} (d_{L,k}),  \tilde I_{R,k}(d_{R,k})\big ) \\
    &&& ~0\leq p_{1}< p_2 < \cdots < p_{L}, d_{L,k}\geq 0, d_{R,k} \geq 0\\
    &&& \frac{1}{L}\sum_{k=1 }^{L}p_{k} \leq 1.
  \end{aligned}
\end{equation}
Note that the objective of problem \eqref{eq:relaxedproblem} has been
substituted in \eqref{eq:relaxedproblem3} with an expression which is
still non-negative and non-decreasing in $d$ for a fixed $p_k$ and
non-increasing in $p_k$ for a fixed $d$; i.e., all the properties and
arguments that led to Algorithm \ref{algo1} are still valid. Thus, the
approach of solving this problem is similar to the one presented in
Section \ref{subsec:Perfect}, with the only difference being that both
$\tilde I_{L,k} (d_{L,k})$ and $\tilde I_{R,k} (d_{R,k})$ exhibit an
easily-interpretable dependence on $p_k$ and $d_{L,k}, d_{R,k}$. Algorithm \eqref{algo2} contains the simplified
algorithm and Appendix \ref{appB} shows the detailed proof.

\begin{algorithm}[t!]
  \small
  \caption{Constellation design: Perfect knowledge of the first,
    second, fourth moments}
  \begin{algorithmic} \label{algo2} \STATE $~$ \STATE $[t^*,
    \symbolconstellation^*]$ = {\bf Bisection}( ); \STATE
    $~$\line(1,0){450}
  \end{algorithmic}
  \begin{algorithmic}
    \STATE {\bf function}: $\symbolconstellation$ = {\bf
      ConstellationDesign}( t ) \STATE $p_1 = 0;$ $d_{L,1} = \infty;$
    $d_{R,L} = \infty;$ \STATE $d_{R,k} = \sqrt{2ts(0)};$
    \FOR{$k=2,3,\cdots, L$} \STATE Find the smallest $p_{k}>p_{k-1}$
    such that
    $\frac{(p_{k}-p_{k-1}^*)^2}{2\left(\sqrt{s(p_{k})}+\sqrt{s(p_{k-1}^*)}\right)^2}
    =t;$ \IF{$k \neq L$} \STATE $d_{R,k} = \sqrt{2ts(p_{k})};$
    \ENDIF
    \STATE $d_{L,k} = \sqrt{2ts(p_{k})};$ \STATE $\mathcal{I}_k =
    [p_k+\sigma^2-d_{L,k},p_k+\sigma^2+d_{R,k}];$
    \ENDFOR
    \STATE {\bf end} {\bf ConstellationDesign} \STATE $\mathcal{P} =
    [p_1,p_2,\cdots,p_L];$ \RETURN $\symbolconstellation,
    \mathcal{D}$;
  \end{algorithmic}
\end{algorithm}

Note that the suggested design leads to an algorithm which can be
employed in very general channel models even when there is no
knowledge of the underlying channel distribution, and that the
approximation gets better as $L$ increases. This is because, as $L$
increases, the transmitted powers are packed closer together,
the decoding regions (i.e., $\{d_{R,k},d_{L,k}\}$) get smaller, and thus the approximation \eqref{eq:approxI} gets tighter.  Figure \ref{FIgA}-(b) shows this numerically with
$\symbolconstellation_{K,\gamma}$ and
$\hat\symbolconstellation_{K,\gamma}$ as a function of the
constellation size $L$ in Rician fading with $K = 0$ dB and $\gamma =
5$ dB. We see that, with increasing constellation sizes, the
approximation \eqref{eq:approxI} gets tighter, which means that both
designs lead to similar error exponents, and to approximately the same
SER. %Observe also that only for $L=2$ the difference between the two designs is significant.  Lastly, note that the constellation derived in this section does not maximize the exact rate function since we have used a second order approximation of the rate functions to derive it. However, this constellation design, does not assume any specific channel distribution, and uses only the first, second and fourth moments of the channel for the design.

\begin{comment}
  \begin{figure}
    \begin{minipage}{1\textwidth}
      \centering \subfigure[$K=6$ dB, $\gamma= 0$ dB]{
        \epsfig{figure=Fifth_SNR5_K6.pdf,width=7.3cm}} \subfigure[ $K
      = 0$ dB, $\gamma = 5$ dB]{
        \epsfig{figure=Six_SNR5_K0.pdf,width=7.5cm}}
      \caption{(a) Nakagami-m fading channel, (b) Comparison of
        $\symbolconstellation_{K,\gamma}$,
        $\hat\symbolconstellation_{K,\gamma}$ as a function of $L$}
      \label{Fig8}
    \end{minipage}
  \end{figure}
\end{comment}

\subsection{Robust constellation design}
\label{subsec:robust}

Assuming perfect knowledge of the channel distribution or even its
first few moments may not be realistic due to changing propagation
environments associated with user mobility and estimation errors. This
motivates the need for designs which take into account uncertainties
in the channel statistics. We build upon the design principles laid
out in the previous sections to develop a design that performs well
even in the face of channel uncertainties.

To exemplify the suggested approach, we are going to assume that the
terminals have estimated the first few moments of the underlying
channel distribution up to some bounded uncertainty. We want to find a
constellation that could work well for all channels with moments
inside their uncertainty region. Recall that $$\expectation \lsb U_k^2
\rsb = s (p_k) = \alpha_1 p_k^2 + \alpha_2 p_k+ \alpha_3,$$ where $a_2=2\sigma^2$ and $a_3=\sigma^4$. Thus, for
a fixed $p_k$, $\expectation \lsb U_k^2 \rsb $, and hence the rate
function approximation depends on the channel and noise statistics
only through $\alpha_1$ and $\sigma$. We then define the following set
$$\mathcal{F} = \{(\alpha_1 , \sigma): \alpha_{\min}< \alpha_1 <
\alpha_{\max}, \sigma_{\min} < \sigma < \sigma_{\max} \},$$ and note
that for each $f = (\tilde{\alpha}_1, \tilde{\sigma}) \in
\mathcal{F},$ we can define $$s_f(p) \triangleq \tilde{\alpha}_1 p^2 +
\tilde{\alpha}_2 p + \tilde{\alpha}_3,$$ where $\tilde{\alpha}_2 = 2
\tilde{\sigma}^2, \tilde{\alpha}_3 = \tilde{\sigma}^4$.

Then, in order to maximize the approximate worst-case rate function
for all possible channels inside the uncertainty region, we modify
problem \eqref{eq:relaxedproblem3} in the following way:
\begin{equation}
  \begin{aligned}
    \label{eq:relaxedproblem3a}
    \negmedspace
    & \underset{ \{p_k , d_{L,k} , d_{R,k}\}_{{k} \in [L] }}{\text{maximize}} & & \min_{f \in \mathcal{F},k \in [L]} \Bigg( \frac{d_{L,k}^2}{s_f(p_k)}, \frac{d_{R,k}^2}{s_f(p_k)}\Bigg ) \\
    &&& ~0\leq p_{1}< p_2 < \cdots < p_{L}, d_{L,k}\geq 0, d_{R,k} \geq 0\\
    &&& \frac{1}{L}\sum_{k=1 }^{L}p_{k} \leq 1.
  \end{aligned}
\end{equation}
In Appendix \ref{appC} we show how to solve problem
\eqref{eq:relaxedproblem3a} and design a constellation which maximizes
the error exponent for all statistics in $\mathcal{F}$. The main
difference between this design as compared to the previous two
algorithms is the need for using power levels and decoding regions
which would work well for any channel statistics inside the bounded
uncertainty of the channel's moments. To satisfy this, the consecutive
power levels and decoding regions are generally spread out as far
apart as the worst channel requires.  Note also that if there is no
uncertainty, Algorithm \ref{algo3} reduces to Algorithm
\ref{algo2}. An important aspect of this approach is that problem
\eqref{eq:relaxedproblem3a}, in contrast to the problems
\eqref{eq:relaxedproblem} and \eqref{eq:relaxedproblem2}, may be
infeasible.  Such an example is presented below.

\begin{algorithm}[t!]
  \small
  \caption{Constellation design: Robust constellation design}
  \begin{algorithmic} \label{algo3} \STATE $~$ \STATE $[t^*,
    \symbolconstellation^*]$ = {\bf Bisection}( ); \STATE
    $~$\line(1,0){450}
  \end{algorithmic}
  \begin{algorithmic}
    \STATE {\bf function}: $\symbolconstellation$ = {\bf
      ConstellationDesign}( t ) \STATE $p_1 = 0;$ $c_0 = -\infty$;
    $c_L = \infty$; \FOR{$k=1,2,\cdots, L-1$} \STATE $c_k = \sup_{f
      \in \mathcal{F}}\left(\sigma^2+ t\sqrt{s_f(p_k)}\right)+p_k;$
    \STATE Find the smallest $p_{k+1}>p_{k}$ such that $p_{k+1}-c_{k}
    -\sup_{f \in \mathcal{F}}\left
      (t^*\sqrt{s_f(p_{k+1})}-\sigma^2\right ) \geq 0;$ \STATE
    $\mathcal{I}_k = [c_{k-1},c_{k}];$
    \ENDFOR
    \STATE $\mathcal{I}_L = [c_{L-1},c_{L}];$ \STATE $\mathcal{P} =
    [p_1,p_2,\cdots,p_L];$ \RETURN $\symbolconstellation =
    [\mathcal{P},\mathcal{I}_1,\cdots, \mathcal{I}_L]$; \STATE {\bf
      end} {\bf ConstellationDesign}
  \end{algorithmic}
\end{algorithm}
% Another interesting comment regarding problem
% \eqref{eq:relaxedproblem3a} is that, depending on $\mathcal{F}$,
% given a specific channel realization, different constellation points
% could experience different rate functions. Even though this might be
% undesirable, the above construction, when feasible, guarantees that
% all symbols will experience a approximate error exponent for any
% channel with long-term statistics inside the uncertainty region
% $\mathcal{F}$.

\subsubsection{Infeasibility of the robust constellation design
  problem }
\label{subsec:infeasible}

In this section we present a simple example that shows that, for a
fixed average power constraint $P$, a very high uncertainty on the
channel statistics could lead to infeasibility in the robust
constellation design problem (Section \ref{subsec:robust}). Consider
the case of constructing a constellation with $L=2$, an uncertainty
region $\sigma^2 \in (\epsilon, \frac{1}{\epsilon})$ for some
$\epsilon>0$ and perfectly known $\alpha_1=1$ for simplicity (the case
of Rayleigh fading). Fix $t^*>0$. Then, based on Algorithm \ref{algo3}
we choose $p_1=0$ and
$c_1=\frac{1}{\epsilon}+\frac{t^*}{\epsilon}$. We next choose $p_2$ to
be the smallest $p>0$ that satisfies \small
\begin{align*}
  \negmedspace p-\frac{1+t^*}{\epsilon}-\sup_{\sigma^2 \in
    (\epsilon,\frac{1}{\epsilon})} \left(
    t^*\sqrt{p^2+2\sigma^2p+\sigma^4}-\sigma^2 \right ) \geq 0
  \Leftrightarrow p-\frac{1+t^*}{\epsilon}- \sup_{\sigma^2 \in
    (\epsilon,\frac{1}{\epsilon})}\left(t^*p+\sigma^2(t^*-1)\right)
  \geq 0.
\end{align*}
\normalsize If $t^*\geq1$, then $p\geq t^*p+2\frac{t^*}{\epsilon}$
which is impossible, and if $t^*< 1$, then $p \geq
\frac{1+t^*}{1-t^*}\frac{1}{\epsilon}-\epsilon$. Thus, the smallest
choice of $p_2$ that can be chosen is $p_2=
\frac{1+t^*}{1-t^*}\frac{1}{\epsilon}-\epsilon.$ In this case,
since $$p_1+p_2 \leq 2P \Rightarrow
\frac{1+t^*}{1-t^*}\frac{1}{\epsilon}-\epsilon < 2P,$$ it follows
that, no matter how small $t^*$ is, if the uncertainty is so large
such that $2P<\frac{1}{\epsilon}-\epsilon$, the robust design problem
will be
infeasible. %In other words, given a fixed average power constraint, it is not possible to guarantee any nonzero error exponent if the uncertainty is too large.

\section{Numerical examples}
\label{sec:numerical}

% For demonstration purposes we use a channel matrix $\mathbf{h}$ such
% that each $h_i$ follows the Gaussian distribution
% $\symbolconstellation N \lb \mu , \frac{1}{1+K}\rb$, i.e., we assume
% Rician fading, with $|\mu|^2 = \frac{K}{1+K}, ~\sigma^2_{h} =
% \frac{1}{1+K}.$

This section contains simulation studies which demonstrate and compare
the performance of all the constellation designs proposed in this
work.  In the following sections, for the constellation designs which
depend on an underlying channel, i.e.,
$\symbolconstellation_{K,\gamma}, \hat\symbolconstellation_{K,\gamma},
\hat\symbolconstellation^{(a)}_{K,\gamma}$, the $(K,\gamma)$ channel
is referred to as the \textit{nominal} channel, whereas any other
channel is referred to as a \textit{mismatched} channel.

\subsection{Comparison with a pilot-based system with PAM and a
  noncoherent system with ASK}

Consider a block-fading Rician fading channel $h_{i} \sim
\mathcal{CN}(\mu,\sigma^2)$ with coherence time $T$ and with $n$
antennas at the receiver. We assume that both the transmitter and the
receiver know the channel statistics but not the exact channel
realization.

In the first numerical example we compare the proposed
$\symbolconstellation_{K,\gamma}$ with the ASK constellation, and with
a system that uses a PAM constellation (referred to as \textit{PAM
  system}) assuming a binary reflected Gray Code (BRGC)
\cite{goldsmith2005wireless}. In the PAM system the transmitter uses
the first $T_{l}$ slots of each coherence interval to transmit pilot
symbols. Based on the received signals in these slots, the receiver
derives the MMSE channel estimates $\{\hat h_i\}$ at the end of the
$T_l$ learning slots. Using these estimates it decodes the symbols
transmitted during the remaining $T-T_{l}$ slots of the coherence
interval. Note that, assuming a constellation size of $L$, the
effective rate of such a system is $$\frac{T-T_l}{T}\log_2(L).$$ The
noncoherent system that uses ASK i.e., amplitudes that are equally
spaced apart, performs decoding using an energy-based ML receiver.

\begin{figure}
  \begin{minipage}{1\textwidth}
    \centering \subfigure[$K=-\infty$ dB, $\gamma=10$
    dB]{\epsfig{figure=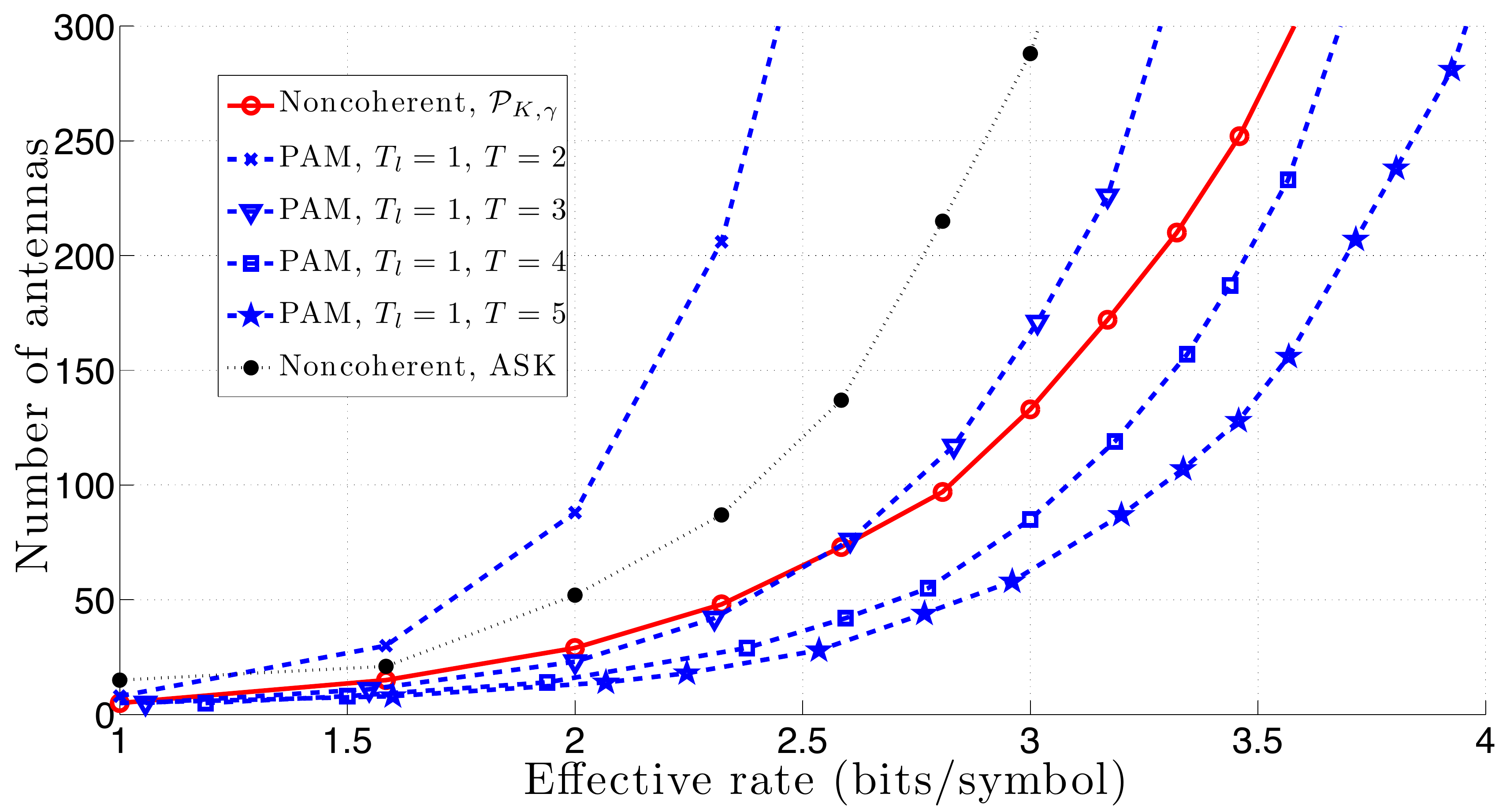,width=7.6cm}} \qquad
    \subfigure[$K=0$ dB, $\gamma=10$ dB
    ]{\epsfig{figure=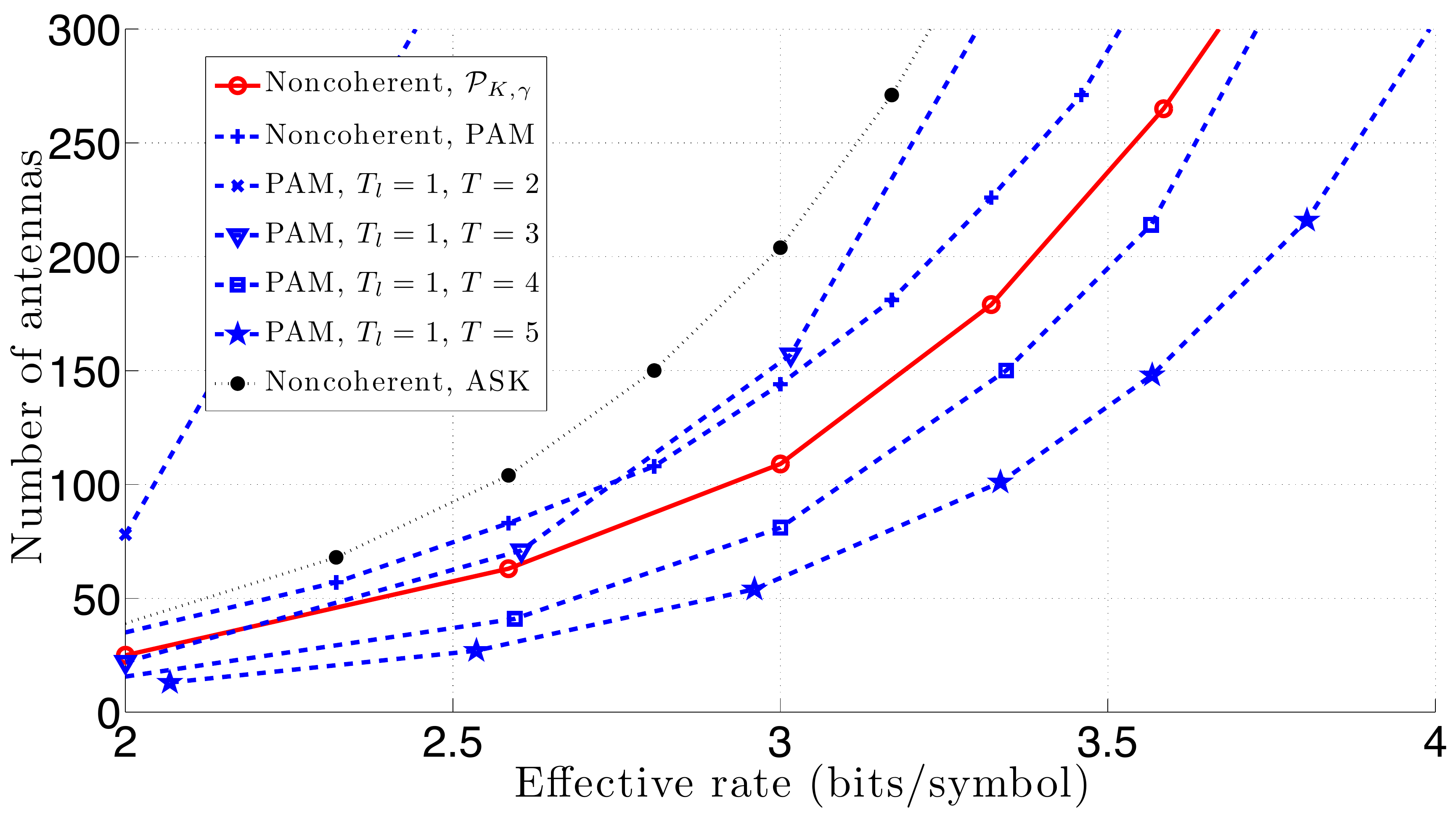,width=7.4cm}} \qquad
    \subfigure[$K=-5$ dB, $\gamma=10$
    dB]{\epsfig{figure=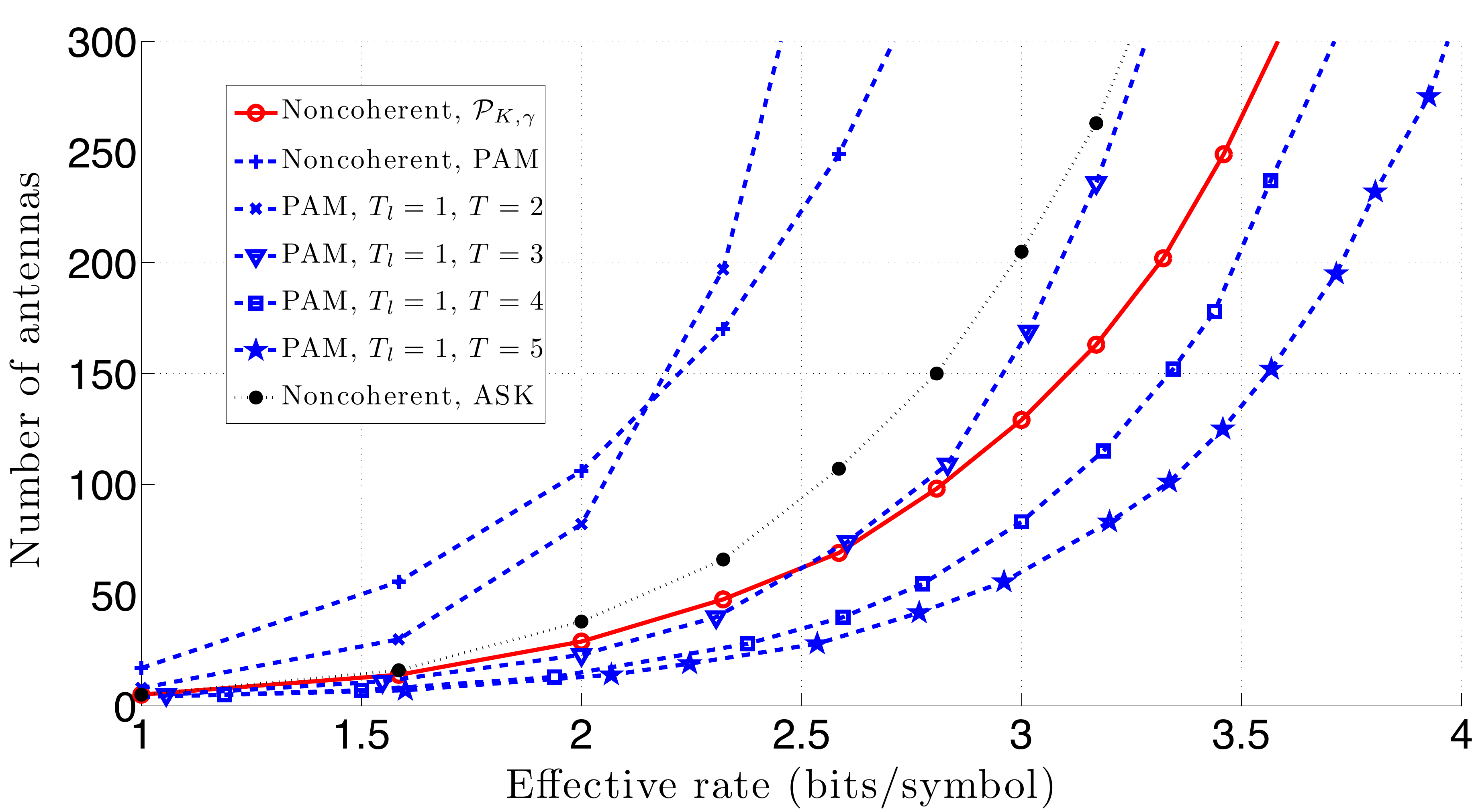,width=7.5cm}} \qquad
    \subfigure[$L=4$]{\epsfig{figure=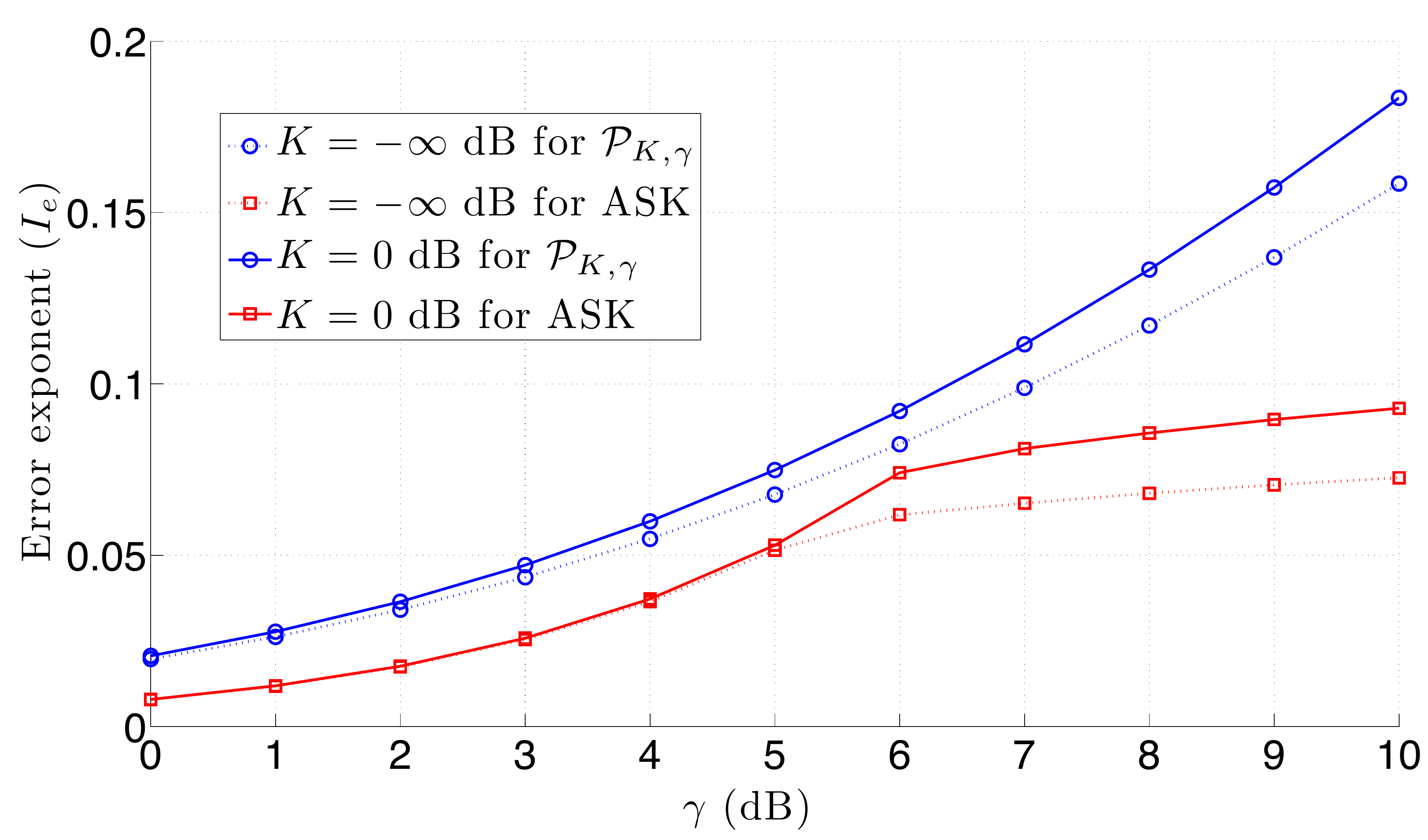,width=7.1cm}}
    \caption{(a)-(c) Minimum $n$ for a target BER $= 10^{-3}$, (d)
      Error exponents of ASK and $\symbolconstellation_{K,\gamma}$}
    \label{Fig7d} \end{minipage}
\end{figure}

Figures \ref{Fig7d}-(a) to \ref{Fig7d}-(c) plot the minimum number of
antennas needed to achieve an uncoded $\mbox{BER }=10^{-3}$ for
different $K$ and $\gamma=10$ dB for different coherence times $T$. We
make the following observations:

\begin{itemize}
\item First, our noncoherent constellation design performs
  significantly better than the system with an ASK constellation. For
  example, in Rayleigh fading our constellation needs approximately
  half the number of antennas to achieve the same BER performance.
\item Second, even in the case of high $K$
  (\ref{Fig7d}-(b),\ref{Fig7d}-(c)), which is known to all the
  receivers, our system performs better than the noncoherent PAM
  system (i.e., $T_l=0$) which exploits the phase of the LOS component
  of the channel. Note that this is not the case with our system which
  only uses the $K-$factor to decode the symbols.
\end{itemize} 

Note that Figure \ref{Fig7d}-(a) does not show the performance of the PAM system since in Rayleigh fading the latter cannot reach the BER target for any number of antennas as the phase of the transmitted symbol is
  completely destroyed. We also observe that, for short coherence
times, the proposed constellation design still requires a smaller
number of antennas to reach the BER threshold compared to the PAM
system with $T_l=1$. On the other hand, for higher coherence times,
the PAM system achieves better performance since the gains of learning
are more than the corresponding decrease in the effective rate. Yet,
observe that for small effective rates, e.g., $1-2$ bits/symbol, the
additional number of antennas needed by the energy-based system to
achieve the same BER as PAM is not more than $20$. This shows that
even a simple energy-based architecture design at the receiver, which
requires only envelope detectors, could be enough to transmit
information as reliably as a typical pilot-based system, especially in
channels with small coherence times and high LOS, without the need for
significantly more
antennas. %For example, this system could potentially perform well in a wide-band multi-carrier setting, in which in every coherence bandwidth, the transmitter sends low-rate independent symbols, trading-off the need of complex receiver architectures with spectral efficiency.

Figure \ref{Fig7d}-(d) plots the error exponent $I_e$ for different
values of $\gamma$, and Rician channels with $K=\{\infty,0\}$ dB and
$L=4$, for two noncoherent systems that use ASK and
$\symbolconstellation_{K,\gamma}$. We observe that for all channel
conditions, our noncoherent constellation design achieves a much
higher $I_e$ than the ASK constellation. Also, for high SNR, the error
exponent that uses the ASK constellation is not increasing as fast as
the system with $\symbolconstellation_{K,\gamma}$. This is due to the
fact that the power levels are fixed, and do not adapt to the channel
conditions. This is not the case with the
$\symbolconstellation_{K,\gamma}$ constellation.

\subsection{SER performance comparison of
  $\symbolconstellation_{K,\gamma}$,
  $\hat\symbolconstellation_{K,\gamma}$, $\symbolconstellation^{min}$,
  ASK}

\begin{figure}
  \begin{minipage}{1\textwidth}
    \centering
    % \subfigure[$K=-\infty$ dB, $\gamma=0$ dB]{
    % \epsfig{figure=First_Experiment_-50_0.pdf,width=7.6cm}} \qquad
    % \subfigure[$K=0$ dB, $\gamma=0$ dB]{
    % \epsfig{figure=First_Experiment_-50_0.pdf,width=7.6cm}} \qquad
    \subfigure[$K=-\infty$ dB, $\gamma=5$ dB]{
      \epsfig{figure=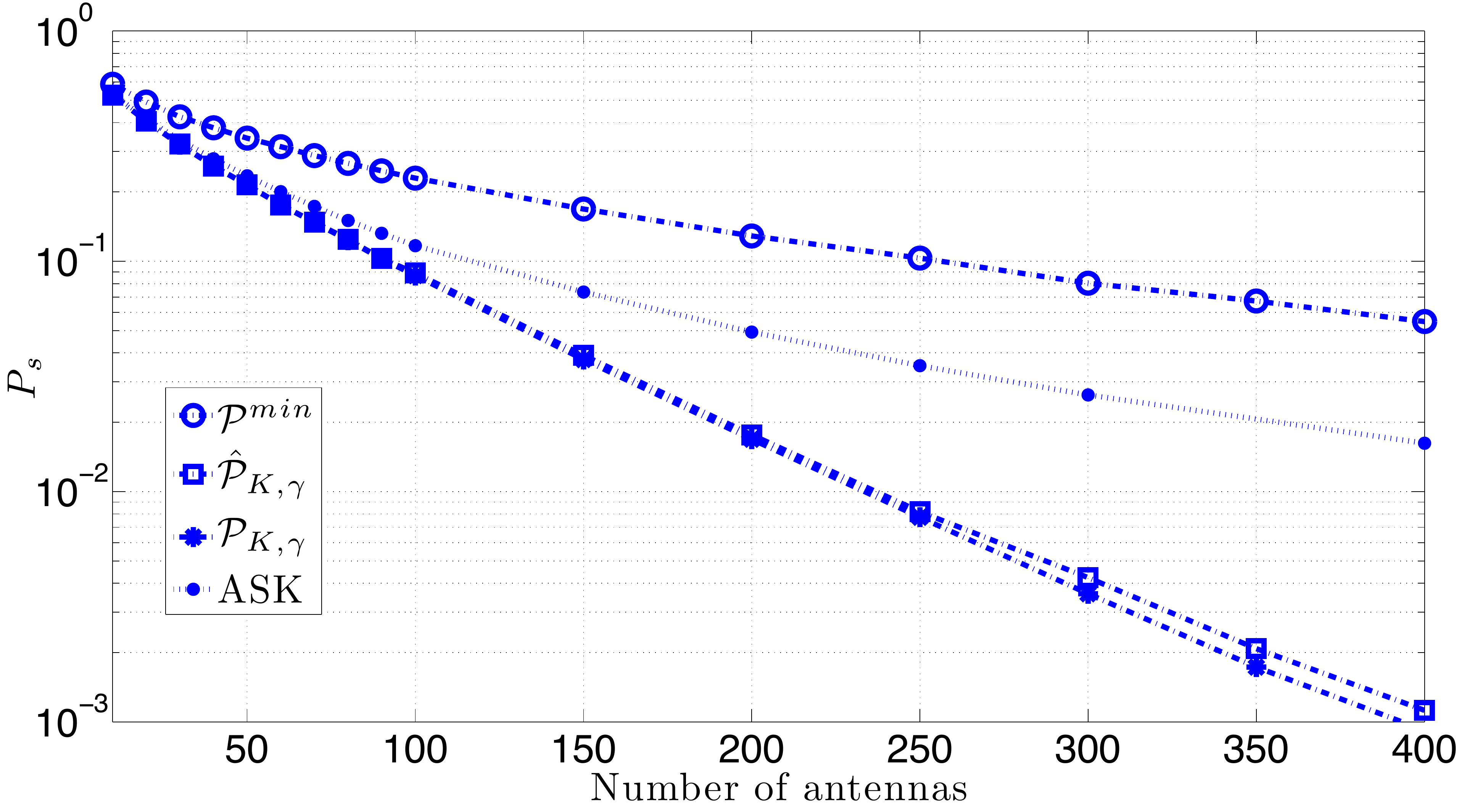,width=8.6cm}} \qquad
    % \subfigure[$K=0$ dB, $\gamma=5$ dB]{
    % \epsfig{figure=First_Experiment_0_5.pdf,width=7cm}} \qquad
    \subfigure[$K=-\infty$ dB, $\gamma=10$ dB]{
      \epsfig{figure=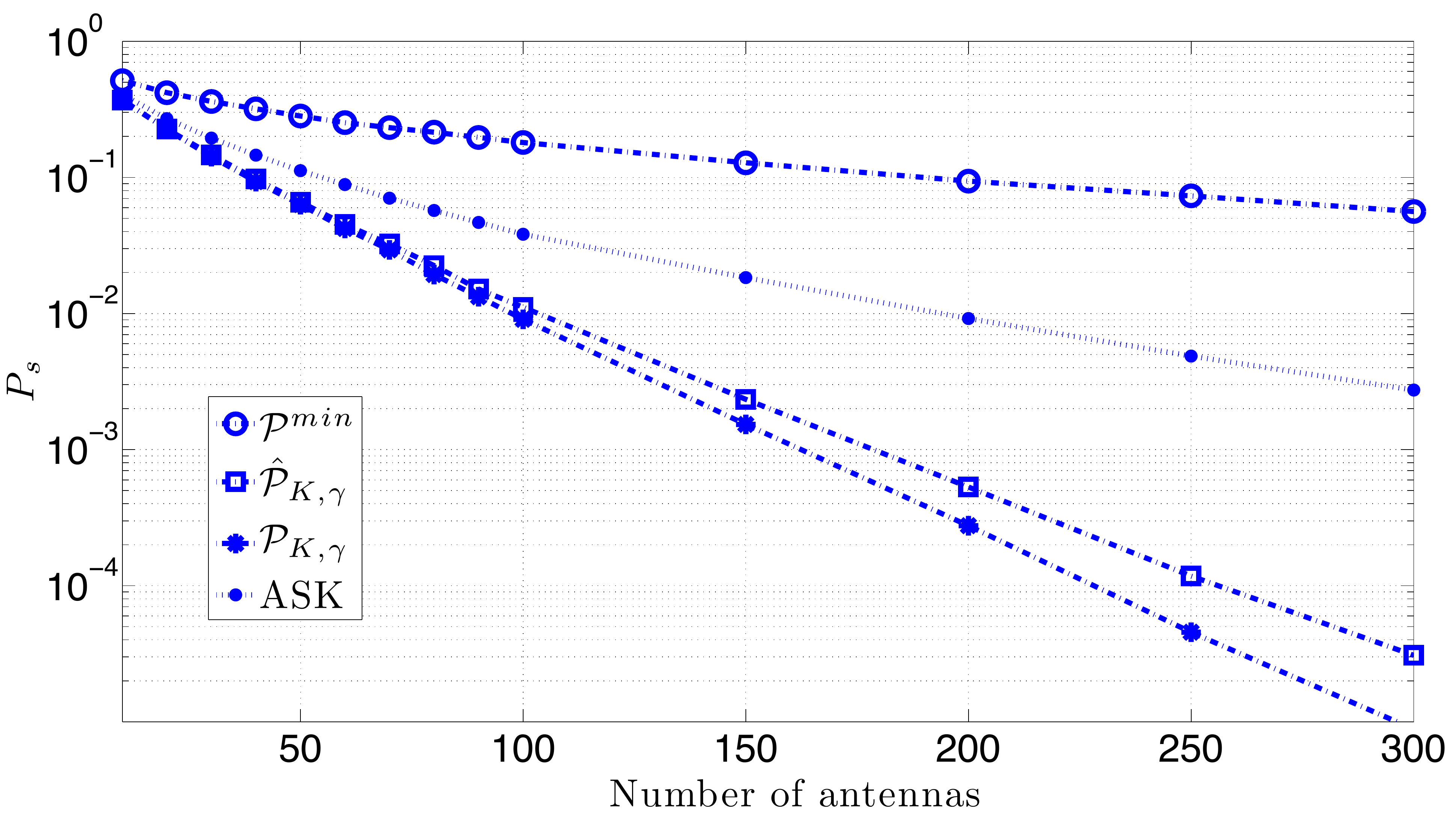,width=8.6cm}} \qquad
    % \subfigure[$K=0$ dB, $\gamma=10$ dB]{
    % \epsfig{figure=First_Experiment_0_10.pdf,width=7cm}} \qquad
    \caption{SER performance comparison of
      $\symbolconstellation_{K,\gamma}$,
      $\hat\symbolconstellation_{K,\gamma}$,
      $\symbolconstellation^{min}$, ASK.}
    \label{Fig6} \end{minipage}
\end{figure}

In the second numerical example (Figure \ref{Fig6}) we present, as a
function of $n$, a numerical SER estimate for a $3-$bit constellation
($L=8$) for channels with $K=-\infty$ dB, i.e., Rayleigh fading and
$\gamma = \{5,10\}$ dB. We consider the following constellations:
$\symbolconstellation_{K,\gamma}$,
$\hat\symbolconstellation_{K,\gamma}$, $\symbolconstellation^{min}$
and ASK. As expected, $\symbolconstellation_{K,\gamma}$ achieves
better SER performance than all the remaining designs. Yet, the
difference of the approximate design
$\hat\symbolconstellation_{K,\gamma}$ from
$\symbolconstellation_{K,\gamma}$ is not significant, especially at
low SNR. Also, the minimum distance design
$\symbolconstellation^{min}$ is significantly worse than any other
design, except for very low SNRs, where the gap in the performance is
smaller. %Similar performance curves can be derived for Rician fading, but are omitted due to space constraints.

\begin{figure}
  \begin{minipage}{1\textwidth}
    \centering \subfigure[$K=-11$ dB, $\gamma=9$ dB]{
      \epsfig{figure=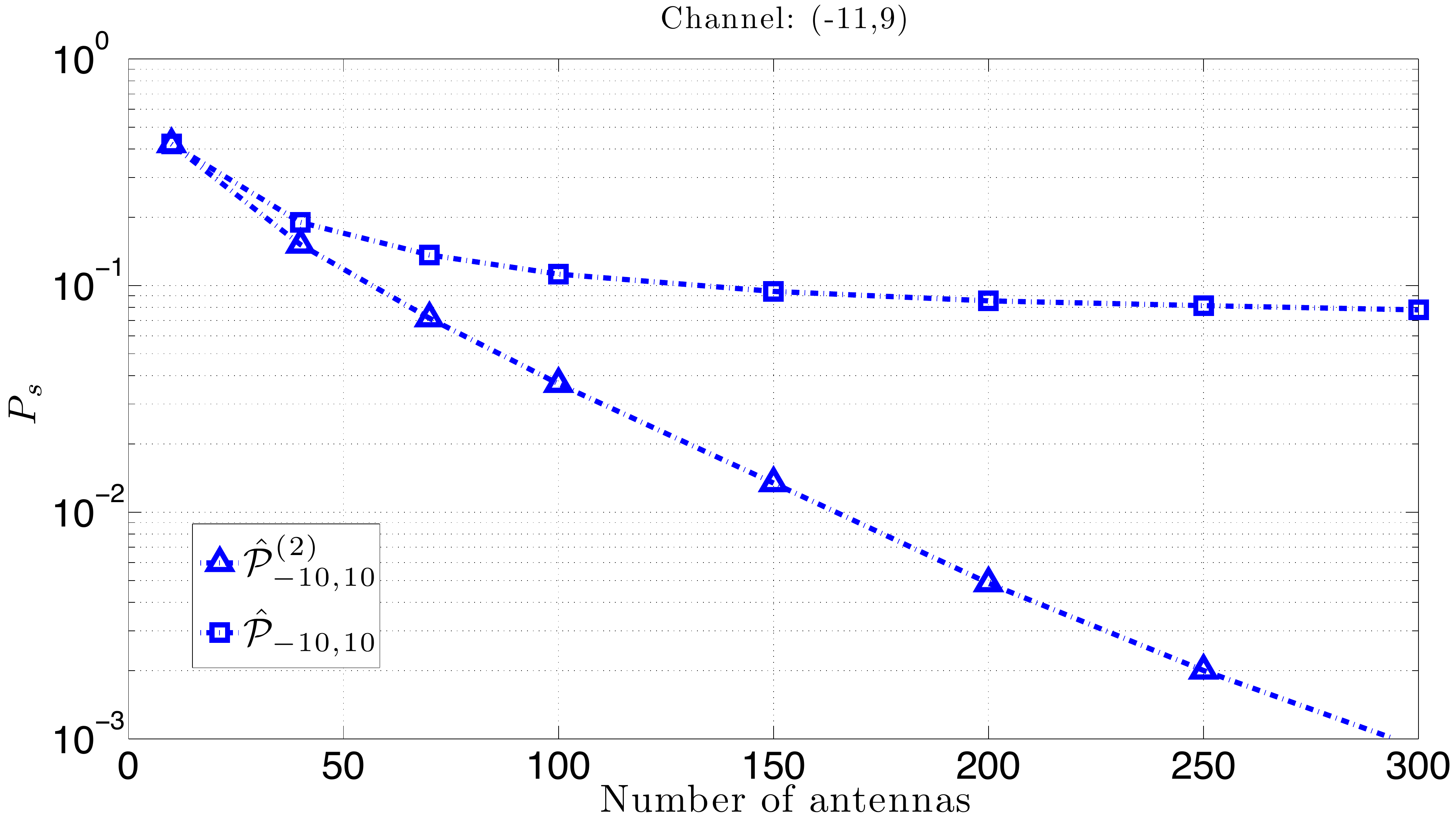,width=7.5cm}}
    \qquad \subfigure[$K=-9$ dB, $\gamma=9$ dB]{
      \epsfig{figure=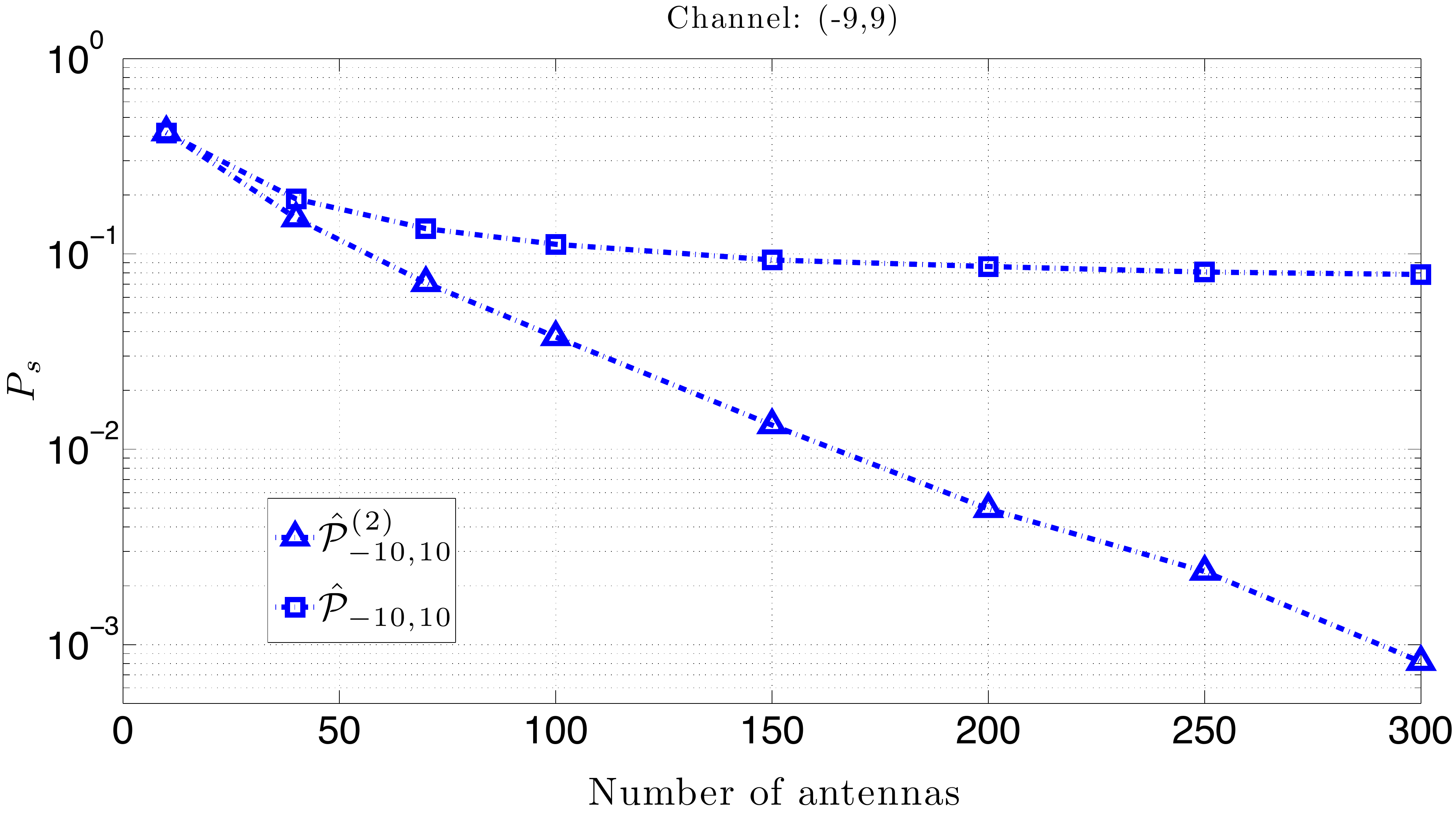,width=7.6cm}}
    \qquad \subfigure[$K=-9$ dB, $\gamma=11$ dB]{
      \epsfig{figure=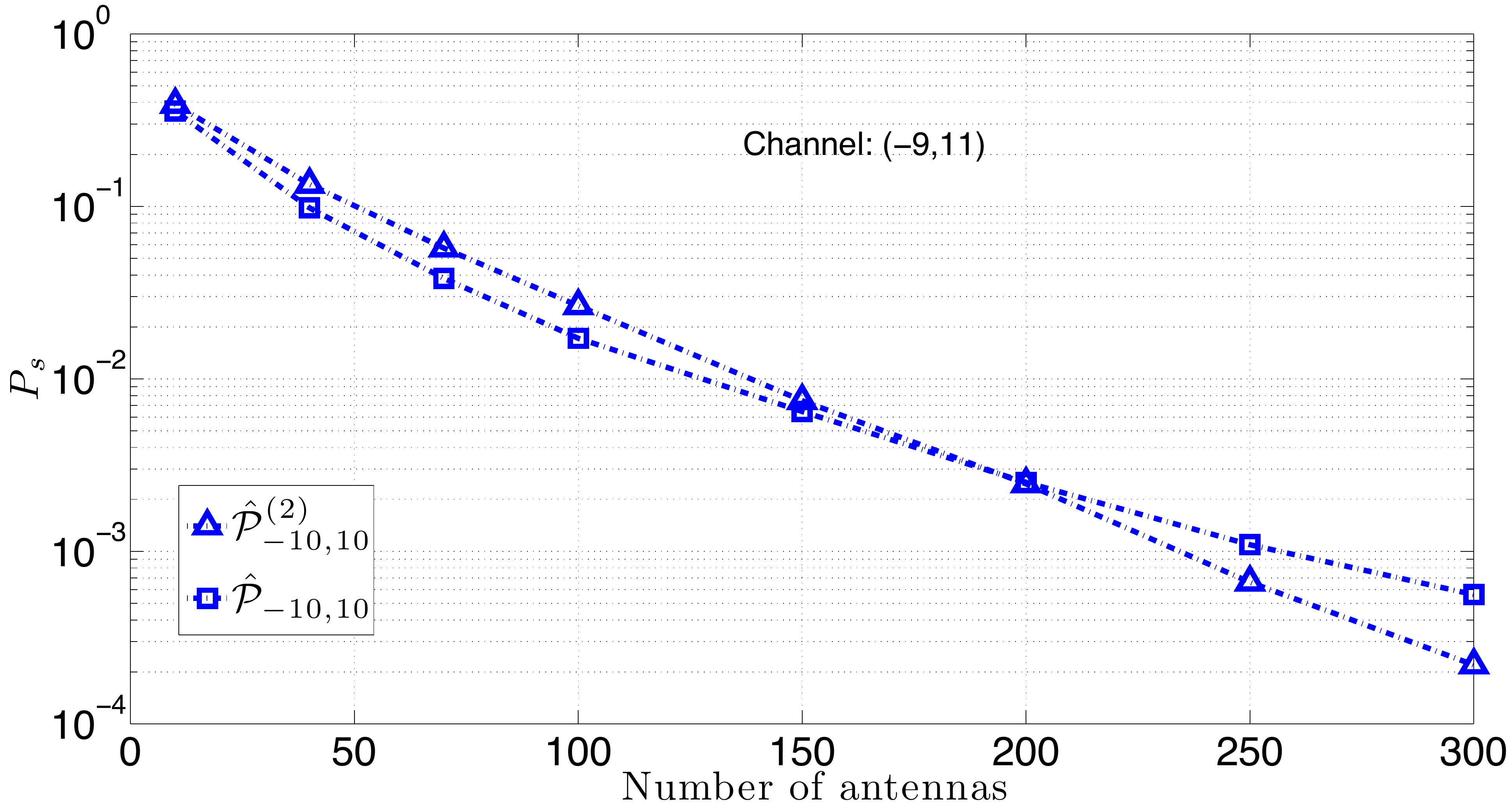,width=7.5cm}}
    \qquad \subfigure[$K=-11$ dB, $\gamma=11$ dB]{
      \epsfig{figure=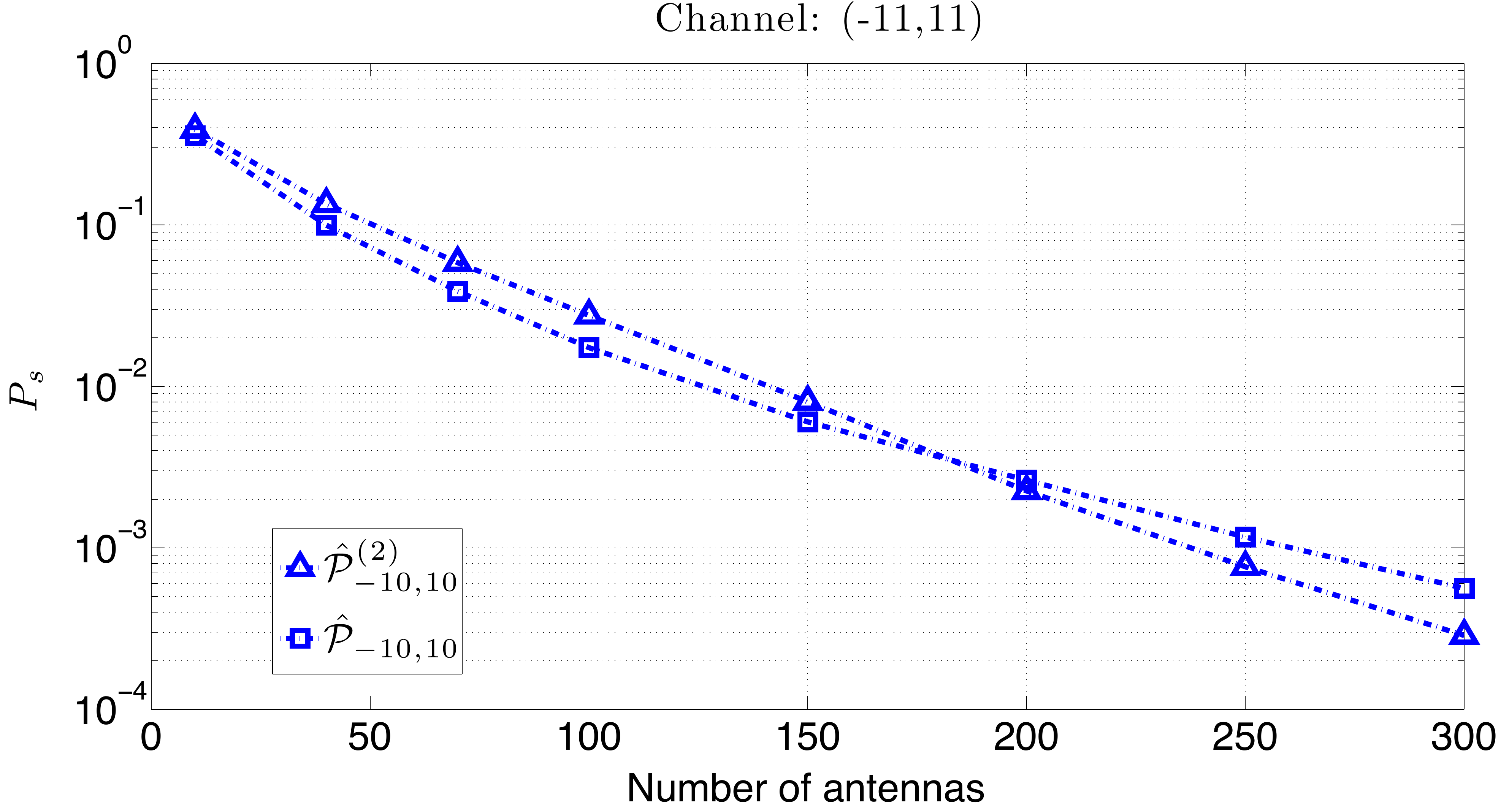,width=7.5cm}}
    \qquad
    \caption{SER performance of the robust constellation designs in
      mismatched channel}
    \label{Fig7} \end{minipage}
\end{figure}

\subsection{Performance of the robust constellation designs on the
  nominal and mismatched channels}

% \subsubsection{SER performance of the robust constellation designs
% in mismatched channel}

In the third numerical example we demonstrate the inefficiency of the
$\hat\symbolconstellation_{K,\gamma}$ constellation in a mismatched
channel and the ability of $\hat\symbolconstellation^{(a)}_{K,\gamma}$
to sustain good performance. Specifically, we consider $L=8$, the case
of a user with $2-$dB of uncertainty in both $K$ and $\gamma$ values
and that the center of the uncertainty interval corresponds to the
$(-10,10)$ channel. This channel is approximately Rayleigh fading ($K$
is very low) with a high SNR value. In Figures  \ref{Fig7}-(a),
\ref{Fig7}-(b), \ref{Fig7}-(c), \ref{Fig7}-(d) we plot the Monte Carlo
SER estimate of the $\hat\symbolconstellation_{-10,10}$ and
$\hat\symbolconstellation^{(2)}_{-10,10}$ designs on the
$\{(-9,9),(-9,11),(-11,9),(-11,11)\}$ channels respectively. Observe
the huge performance loss that occurs due to the overestimation of the
SNR. Smaller performance loss is observed due to the uncertainty on
the value of $K$, or when the SNR is
underestimated. %Comparative figures between $\symbolconstellation^{(1,2)}_{-10,10}$ and $\symbolconstellation^{(1)}_{-10,10}$ results in very similar observations and are omitted due to space considerations.
\begin{comment}
  Similarly, Figure \ref{Fig7}-(e), \ref{Fig7}-(f) plots the same
  quantity for the $\symbolconstellation^{(1,2)}_{-10,10}$ and
  $\symbolconstellation^{(1)}_{-10,10}$ designs. We observe again that
  when the SNR is overestimated, the usage of the
  $\symbolconstellation^{(1)}_{-10,10}$ design could have a
  detrimental effect on the performance. Observe that this is not the
  case with the $\symbolconstellation^{(1,2)}_{-10,10}$ which takes
  into account the bounded uncertainty on the channel statistics.
\end{comment}

\begin{figure}
  \begin{minipage}{1\textwidth}
    \centering
    % \subfigure[$K=-3$ dB, $\gamma=3$ dB ]{
    % \epsfig{figure=Third_Experiment_D-3_3_C_-3_3_Approx.pdf,width=7cm}}
    % \qquad \subfigure[$K=-2$ dB, $\gamma=2$ dB ]{
    % \epsfig{figure=Sec_Experiment_D-3_3_C_-2_2_Approx.pdf,width=7cm}}
    \subfigure[$K=-10$ dB, $\gamma=10$ dB]{
      \epsfig{figure=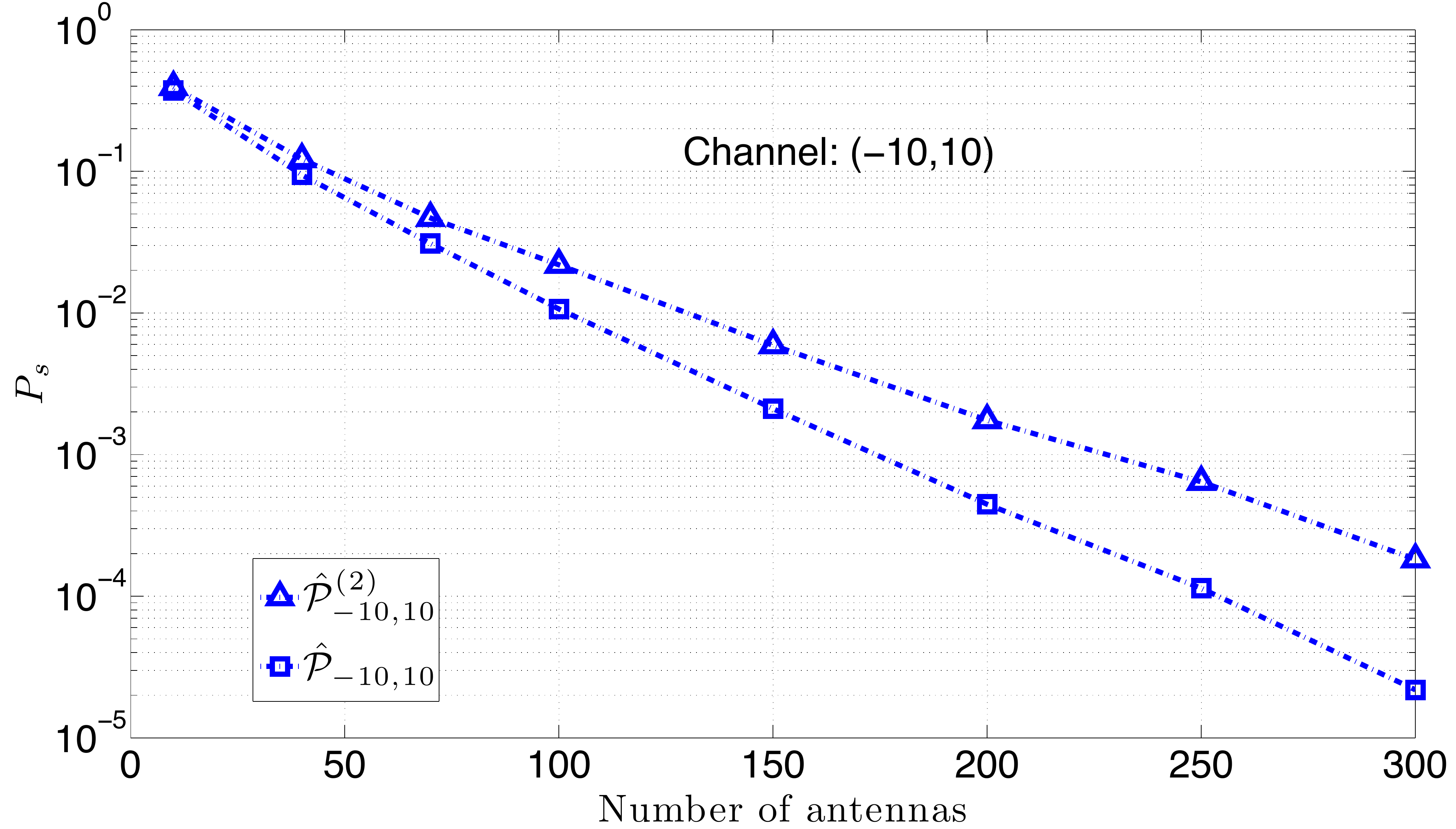,width=7.8cm}}
    \qquad \subfigure[$K=6$ dB, $\gamma= 0$ dB]{
      \epsfig{figure=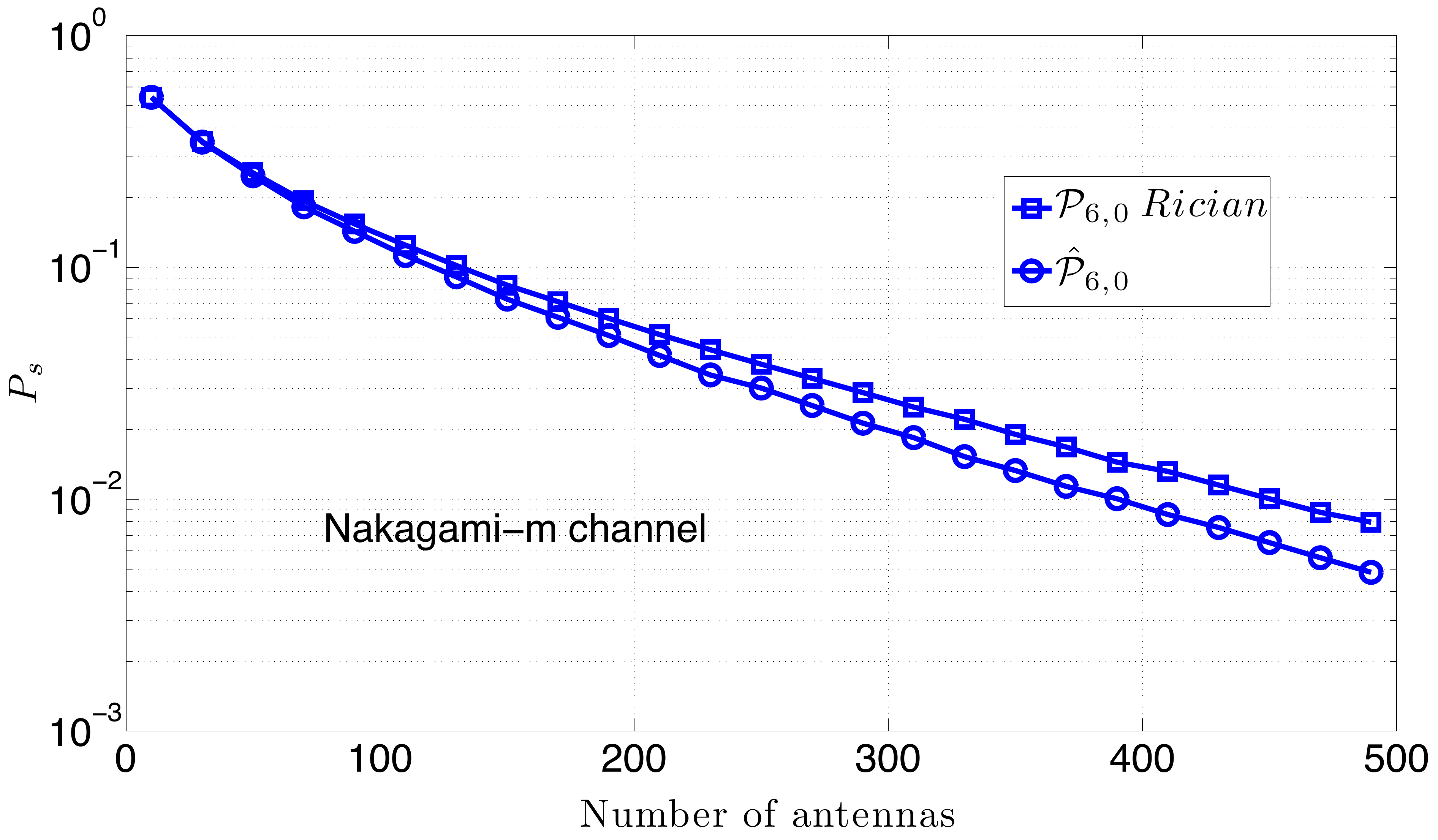,width=7.8cm}}
    \caption{(a) SER performance of the robust constellation designs
      in nominal channel (b) Nakagami-m fading channel}
    \label{Fig7b} \end{minipage}
\end{figure}

% \subsubsection{SER performance of the robust constellation designs
% in nominal channel}

Figure \ref{Fig7b}-(a) presents the SER performance of
$\hat\symbolconstellation^{(2)}_{-10,10}$ and
$\hat\symbolconstellation_{-10,10}$ in the $(-10,10)$ channel to show
that even with nominal statistics, the performance of the robust
design is close to that of the design that is explicitly optimized for
the nominal statistics. This shows that the maximum performance loss
due to the robust design compared to a constellation optimized for a
known channel is tolerable, especially considering the fact that not
taking into account the uncertainty could lead to a significant
performance deterioration as presented in the previous numerical
example.

% Also, note that for channels with low SNR, the difference between
% $\symbolconstellation^{(2,2)}_{K,\gamma}$ and
% $\hat\symbolconstellation_{K,\gamma}$ is trivial (e.g.,
% Figure \ref{Fig7b}-(a)), whereas
% $\symbolconstellation%^{(2,2)}_{K,\gamma}$ can still provide good performance in mismatched channels (e.g., Figure \ref{Fig7b}-(b)).
% \qquad\subfigure[$K=0$ dB, $\gamma=10$ dB]{
% \epsfig{figure=forth_K0_SNR10,width=7cm}} \qquad

\subsection{Performance on a Nakagami-$m$ fading channel}

We now show an example in which using
$\symbolconstellation_{K,\gamma}$ designed for a Rician fading channel
leads to a worse performance compared to a
$\hat\symbolconstellation_{K,\gamma}$ in a Nakagami-$m$ fading
channel. This shows that not taking into account the uncertainty in
the channel distribution, and over-optimizing the constellations for
the Rician channel, could lead to worse performance than a much
simpler constellation design which is based only on the first four
moments of the channel. Specifically, consider the case of a channel
for which it holds that $$\expectation[h_i]
=\sqrt{\frac{K}{K+1}},~E\left[|h_i|^2\right]=1,
E\left[\left|h_i-\sqrt{\frac{K}{1+K}}\right|^2\right] =
\frac{1}{1+K}.$$ This channel could correspond to a Rician channel,
i.e., $h_i \sim \mathcal{CN}\lb
\sqrt{\frac{K}{K+1}},\frac{1}{1+K}\rb,$ or a Nakagami-m channel with
$\Omega=1$ and $m$ such that
$\frac{\Gamma(m+\frac{1}{2})}{\Gamma(m)}\frac{1}{\sqrt{m}} =
\sqrt{\frac{K}{K+1}}.$ Figure \ref{Fig7b}-(d) plots $P_s$ for $\gamma =
0$ dB and $K = 6$ dB in a Nakagami-$m$ fading channel using $L=8$, for
the following two scenarios:

\begin{enumerate}
\item a Rician fading channel model and the
  $\symbolconstellation_{6,0}$ constellation design
\item Only the first four moments are perfectly estimated and used in
  the $\hat\symbolconstellation_{6,0}$ design.
\end{enumerate}
Observe that assuming Rician fading and using the corresponding
constellation, leads to a worse performance in Nakagami-$m$ fading
than using a constellation design which takes into account only the
first few moments of the channel distribution.

  \begin{comment}
    since in that case
    \begin{align}
      \expectation[h_i]
      &=\sqrt{\frac{\Omega}{m}}\frac{\Gamma(m+\frac{1}{2})}{\Gamma(m)}
      =
      \sqrt{\frac{K}{K+1}}, \\
      E\left[\left|h_i-\sqrt{\frac{K}{1+K}}\right|^2\right] &=
      \Omega\left(1-\left(\frac{\Gamma(m+\frac{1}{2})}{\Gamma(m)}
        \right)^2\frac{1}{m}\right) = \frac{1}{1+K}.
    \end{align}
  \end{comment}

  \begin{comment}
    \begin{figure}
      \begin{minipage}{1\textwidth}
        \subfigure[ $K = 0$ dB, $\gamma = 5$ dB]{
          \epsfig{figure=Six_SNR5_K0.pdf,width=7.5cm}}
        \caption{(a) , (b) Comparison of
          $\symbolconstellation_{K,\gamma}$,
          $\hat\symbolconstellation_{K,\gamma}$ as a function of $L$}
        \label{Fig8}
      \end{minipage}
    \end{figure}
  \end{comment}

  \section{Conclusions and Future Work}
  \label{sec:conc}
  We have formulated and solved the single-shot constellation design
  problem for a noncoherent SIMO system with a large number of
  antennas and an average energy-detection-based receiver. We present
  asymptotically optimal constellation designs with respect to the
  achieved error exponent when the system has perfect knowledge of the
  channel statistics. Then, we present a constellation design which
  requires only the knowledge of the first four moments of the fading
  statistics.  The gap to optimality with this design becomes smaller
  with larger constellation sizes. Lastly, we present a robust
  counterpart of our designs which takes into account the uncertainty
  in the channel statistics. We exemplify the performance of all the
  proposed constellations, and compare them with existing
  symbol-by-symbol noncoherent schemes in typical scenarios as well as
  in pilot-based schemes. The results show that our designs are better
  than the noncoherent ASK modulation schemes and that they exhibit
  better BER performance than a pilot-based scheme that uses PAM
  modulation when the coherence time is small. The proposed system
  asks for a very simple encoding and decoding and for a receiver
  which only measures the average received energy across all antennas.

  Our findings here suggest that simple receiver architectures are
  promising alternatives to complex coherent designs for the large
  antenna systems of the not-too-distant future.  We did not however
  explore the full range of optimizations that could potentially be
  carried out in such a setup.  We list some directions for future
  research in the following:

  \begin{itemize}
  \item Antenna correlation and how this affects the performance. This
    is especially relevant as antenna form factors go down with
    increasing numbers of antennas.
  \item Constellation designs for a multiuser noncoherent SIMO
    system. Initial results towards this direction appear in
    \cite{manolakos2014}.
  \end{itemize}

  \small \appendices
  % \section{Constellation design: Perfect knowledge of channel
  % distribution}
  \section{}
  \label{appedA}

  To begin, without loss of generality, index the constellation points
  such that $0\leq p_1<p_2<\cdots<p_{L}$. Then, fix any codebook
  $\mathcal{P}$ which satisfies the average power constraint, and
  solve \eqref{eq:relaxedproblem} over only the decoding regions
  $\{\mathcal{I}_i\}$, i.e., over $\{d_{L,k},d_{R,k}\}_{k=1}^{L}$.
  This subproblem can be written as
  \begin{equation}
    \begin{aligned}
      \label{eq:relaxedproblem2}
      & \underset{\{d_{L,k},d_{R,k}\}_{k \in [L]} }{\text{maximize}}
      & & \min_{k \in [L]} \big( I_{L,k} (d_{L,k}),  I_{R,k}(d_{R,k})\big ) \\ &\mbox{subject to}&& d_{L,{k+1}}+d_{R,k} = p_{k+1}-p_k, ~~\forall k \in [L-1], \\
      &&& d_{L,{k}} \geq 0, d_{R,{k}} \geq 0,~ \forall k \in [L], \\
    \end{aligned}
  \end{equation}
  Observe that \eqref{eq:relaxedproblem2} is separable to $L-1$
  optimization problems, one for every $k=[L-1]$, since for each $k$,
  the constraints are separable. To see this, the constraint for $k$
  is $d_{L,{k+1}}+d_{R,k} = p_{k+1}-p_k$ and for $k+1$ is
  $d_{L,{k+2}}+d_{R,k+1} = p_{k+2}-p_{k+1}$; the former is a linear
  constraint between $d_{L,{k+1}},d_{R,k}$ and the latter another
  constraint between $d_{L,{k+2}},d_{R,k+1}$. Each one of the
  resulting subproblems identifies the boundary between the $p_{k}$
  and $p_{k+1}$ constellation point such that the minimum between
  those two is maximum\footnote{assuming $d_{L,1}=\infty$ and
    $d_{R,L}=\infty$.}:
  \begin{equation}
    \begin{aligned}
      \label{eq:relaxedproblem3}
      & \underset{d_{L,k+1},~d_{R,k} }{\text{maximize}}
      & &  \min(I_{R,k}(d_{R,k}),I_{L,k+1}(d_{L,k+1})) \\
      &\mbox{subject to}&& d_{L,{k+1}}+d_{R,k} = p_{k+1}-p_k, \\
      &&& d_{L,k+1} \geq 0, d_{R,{k}} \geq 0.
    \end{aligned}
  \end{equation}
  Each of the above problems is solved for $d_{L,k+1},d_{R,k}$ such
  that
  \begin{align}
    \label{eq:bestequation}
    I_{R,k}(d_{R,k}) = I_{L,k+1}(d_{L,k+1}) \Rightarrow
    I_{R,k}(d_{R,k}) = I_{L,k+1}( p_{k+1}-p_k-d_{R,k}).
  \end{align} Note that such $d_{R,k}$ always exists since, for any
  $p_k < p_{k+1}$, $I_{R,k}(d)$ and $I_{L,k+1}(p_{k+1}-p_k -d)$ are
  increasing and decreasing functions of $d$, respectively, with
  $I_{R,k}(0)=0,I_{L,k+1}(0)=0$ and $I_{L,k+1}(p_{k+1}-p_k)>0,
  ~I_{R,k}(p_{k+1}-p_k)>0.$ In other words, increasing $d_{R,k}$
  increases the right error exponent of the $k^{th}$ power level, but
  decreases the left error exponent of the $(k+1)^{th}$ level, and
  there always exists a $d_{R,k}$ for which both are equal. Therefore,
  for any fixed $\mathcal{P}$, the best decoding regions between any
  consecutive constellation points can be calculated by
  \eqref{eq:bestequation}.

  Then, the following optimization problem finds the optimal
  $\mathcal{P}$:
  \begin{equation}
    \begin{aligned}
      \label{eq:relaxedproblem3}
      & \underset{\{p_k\}_{k \in [L]},\{d_{R,k}\}_{k \in
          [L-1]},t}{\text{maximize}}
      & & t \\
      & \text{subject to}
      & & I_{R,k}(d_{R,k}) = I_{L,k+1}(p_{k+1}-p_k - d_{R,k}),~\forall k \in [L-1], \\
      & & &I_{R,k}(d_{R,k}) \geq t,~\forall k \in [L-1],\\
      & & &\frac{1}{L}\sum_{k=1}^{L}p_k = 1, ~~0\leq p_k< p_{k+1}, ~0<d_{R,k}<p_{k+1}-p_k,~\forall k \in [L-1].\\
    \end{aligned}
  \end{equation}
  The solution of \eqref{eq:relaxedproblem3} corresponds the largest
  $t^*$ such that the following problem is feasible:
  \begin{equation}
    \begin{aligned}
      \label{eq:feasibilityproblem5}
      &\text{find} & & \{p_k\}_{k=1}^{L}, \{d_{R,k}\}_{k=1}^{L-1} \\
      & \text{subject to}
      & & I_{R,k}(d_{R,k}) = I_{L,k+1}(p_{k+1}-p_k - d_{R,k}),~\forall k \in [L-1], \\
      & & &I_{R,k}(d_{R,k}) \geq t^*,~\forall k \in [L-1],\\
      & & &\frac{1}{L}\sum_{k=1}^{L}p_k = 1, ~~0\leq p_k < p_{k+1},~0<d_{R,k}<p_{k+1}-p_k,~\forall k \in [L-1].\\
    \end{aligned}
  \end{equation}
  Observe that for $t^*=0$ the above problem is always feasible since
  $I_{R,k}(d)\geq 0$ and $I_{L,k}(d) \geq 0$. Also observe that for
  $t^* = \infty$ it is infeasible due to the finite power constraint
  and the fact that $I_{R,k}(d),I_{L,k}(d)$ are increasing functions
  of $d$.

  The problem now is to find the largest $t^*$ for which
  \eqref{eq:feasibilityproblem5} is feasible. We are going to describe
  in detail the algorithm that finds whether problem
  \eqref{eq:feasibilityproblem5} has a feasible solution for any fixed
  and finite $t^*>0$. The basic idea of this construction is that, for
  any fixed $t^*$, we should find the constellation with the smallest
  average power constraint, as this is the only constraint that could
  lead to infeasibility of \eqref{eq:feasibilityproblem5}. To start,
  fix $t^*>0$ and choose $p_1^*=0$. Choosing a higher value for $p_1$
  can only make the problem more difficult since $I_{R,1}(d)$ is
  decreasing in $p_1$, and $p_2>p_1$, thus also the rate functions for
  $p_2$ and the rest constellation points will be lower. Then, find
  $d_{R,1}>0$, denoted as $d_{R,1}^*$, such that
  \begin{align}
    \label{eq:eqw}
    I_{R,1}(d_{R,1}) = t^*.
  \end{align}
  The above equation has always only one solution since $I_{R,1}(d)$
  is an increasing function of $d$, $I_{R,1}(0) = 0$ and $t^*>0$. The
  solution of \eqref{eq:eqw} leads to the closest point to the right
  of $p_1^*=0$ which should be used as a boundary point for the first
  constellation point. Using a smaller boundary point would lead to a
  smaller rate function than $t^*$ since $I_{R,1}(d)$ is increasing on
  $d$. Until now we have specified $p_1^*, d_{R,1}^*$. Now, we find
  the smallest $p_2>p_1^*+d_{R,1}^*$, denoted as $p_2^*$, such that
  \begin{align}
    \label{eqw2}
    I_{L,2}(p_{2}-p_1^* - d_{R,1}^*) = t^*.
  \end{align}
  Note that for $p_2=p_1^*+d_{R,1}^*$, $I_{L,2}(0) = 0$. If there is
  no $p_2>p_1^*+d_{R,1}^*$ that solves \eqref{eqw2}, then problem
  \eqref{eq:feasibilityproblem5} is infeasible for this $t^*$ and we
  need to repeat the construction for a larger $t^*$. Note that if
  this is the case, i.e., if $ I_{L,2}(p_{2}-p_1^* - d_{R,1}^*)<t^*$
  for all $p_2>p_1^*+d_{R,1}^*$, then choosing a $\hat d_{R,1}^*
  >d_{R,1}^*$ in the previous step would not have made \eqref{eqw2}
  feasible. This is because, for any fixed $p_2>p_1^*+\hat d_{R,1}^*$,
  \begin{align}
    t^* > I_{L,2}(p_{2}-p_1^* - d_{R,1}^*) > I_{L,2}(p_{2}-p_1^* -
    \hat d_{R,1}^*),
  \end{align}
  since $I_{L,2}(d)$ is an increasing function of $d$. On the other
  hand, if \eqref{eqw2} has a solution, we use $p_2^*$ and $d_{R,1}^*$
  to find $d_{L,2}^*$ by $d_{L,2}^* = p_{2}^*-p_1^*-d^*_{R,1}.$

  We calculate the remaining $\{p_k^*\}_{k=3}^{L}$ and decoding
  regions iteratively by first finding $d_{R,k}^*>0$ that solves
  $I_{R,k}(d_{R,k}) = t^*$, and then finding the smallest
  $p_{k+1}^*>p_k+d_{k,R}^*$ which satisfies $I_{L,k+1}(p_{k+1}^*-p_k^*
  -d_{R,k}^*) = t^*$. By construction, this solution corresponds to
  the constellation points with the minimum sum power which achieves a
  minimum left and right rate functions at each constellation point of
  at least $t^*$.

  In words, this is the case since choosing a smaller $d_{R,k}$ than
  $d_{R,k}^*$, or a smaller $p_{k+1}$ than $p_{k+1}^*$, would lead to
  a smaller right error exponent than $t^*$ for the $k^{th}$ point, or
  a smaller left error exponent than $t^*$ for the $(k+1)^{th}$
  point. Then, if it holds that $\frac{1}{L}\sum_{k=1}^{L}p_k^* \leq
  1,$ problem \eqref{eq:feasibilityproblem5} is feasible for
  $t^*$. Identifying the largest $t^*$ for which
  \eqref{eq:feasibilityproblem5} is feasible solves
  \eqref{eq:relaxedproblem}.

  To efficiently perform this procedure we can employ a simple
  bisection algorithm (Algorithm \ref{algo1}). To see this, observe
  that for a $\tilde t$, such that $t^*<\tilde t$, the corresponding
  constellation design leads to higher (or equal) average transmitted
  power (infinite power if the problem is infeasible). This is true
  because, at each step of the constellation design, finding $\tilde
  d_{R,k}$ that satisfies $I_{R,k}(\tilde d_{R,k}) = \tilde t$ will
  lead to a $\tilde d_{R,k}$ with $\tilde d_{R,k} > d_{R,k}^*$, and
  finding $\tilde p_{k}$ which satisfies $I_{L,k}(\tilde p_{k}-\tilde
  p_{k-1} -\tilde d_{R,k-1}) = \tilde t$ will lead to a $\tilde p_{k}$
  with $\tilde p_{k} > p_k^*$. \qed

\begin{comment}
  Similarly, any constellation design with $\tilde p_1 > 0$ is
  suboptimal. Thus, it is possible to find the largest $t^*$ with a
  bisection method, and get the solution of \eqref{eq:relaxedproblem}
  efficiently.  Denote as $t^*_{max}$ the solution of
  \eqref{eq:relaxedproblem}. Then,
  \begin{align}
    P_e \doteq P_U \doteq \left(2-\frac{2}{L}\right)e^{-nt^*_{max}}.
  \end{align}
\end{comment}

\begin{comment}
  \begin{equation}
    \begin{aligned}
      \label{relaxedproblem2}
      & \underset{\{p_k, d_{L,k},d_{R,k}\}_{k \in [L]}
      }{\text{maximize}}
      & & \min_{k \in [L]}   \lb \frac{d^2_{R,k}}{2s(p_k)}, \frac{d_{L,{k+1}}^2}{2s(p_{k+1})} \rb \\
      &\mbox{subject to}&& d_{L,{k+1}}+d_{R,k} = p_{k+1}-p_k\geq 0, ~~\forall k \in [L-1], \\
      &&& d_{L,{k}} \geq 0, d_{R,{k}} \geq 0, \forall k \in [L].
    \end{aligned}
  \end{equation}
\end{comment}

% \section{Constellation design: Perfect knowledge of the first,
% second, fourth moments}
\section{}
\label{appB}
In this appendix we take into account the approximation of the left
and right rate functions shown in \eqref{eq:approxI} to simplify the
algorithm needed for a constellation design that uses only the first,
second and fourth moments. Equation \eqref{eq:bestequation} can now be
written as follows $$ \frac{d_{L,{k+1}}^2}{2s(p_{k+1})} =
\frac{d_{R,k}^2}{2s(p_{k})}=\frac{(p_{k+1}-p_k)^2}{2\left(\sqrt{s(p_{k+1})}+\sqrt{s(p_k)}\right)^2},$$
which means that the feasibility problem
\eqref{eq:feasibilityproblem5} is simplified to
\begin{equation}
  \begin{aligned}
    \label{feasibilityproblem2}
    &\text{find} & & \{p_k\}_{k=1}^L \\
    & \text{subject to}
    & & p_{k+1}-p_k \geq \sqrt{2t^*}\left(\sqrt{s(p_{k+1})}+\sqrt{s(p_k)}\right) \\
    & & &\frac{1}{L}\sum_{k=1}^{L}p_k \leq 1. \\
  \end{aligned}
\end{equation}
Then the procedure described in Appendix \ref{appedA} is now
simplified to the following: Fix $t^*$ and choose $p_1^*=0$. Then,
iteratively choose the smallest $p_{k+1}>p_{k}^*$ for
$k=1,2\cdots,L-1$, such that
$$\frac{(p_{k+1}-p_k^*)^2}{2\left(\sqrt{s(p_{k+1})}+\sqrt{s(p_k^*)}\right)^2}=t^*.$$ 
If no $p_{k+1}>p_{k}^*$ exists, then problem
\eqref{feasibilityproblem2} is infeasible. \qed
\section{}
\label{appC}

In this appendix, we show the details of the robust constellation
design problem. This problem is simplified if we denote a
constellation using $\{p_k\}_{k=1}^{L}$ and $\{c_k\}_{k=1}^{L-1}$,
where $c_k$ is the boundary of the decoding region between the $p_{k}$
and the $p_{k+1}$ constellation point.  Then, using the approximation
shown in \eqref{eq:approxI}, the problem of maximizing the worst case
approximate rate functions for all the channels inside the uncertainty
region $\mathcal{F}$ is expressed as follows:
\begin{equation}
  \begin{aligned}
    \label{robustproblem}
    & \underset{t,\{p_k\}_{k=1}^{L},\{c_k\}_{k=1}^{L-1}}{\text{maximize}} & & t \\
    & \text{subject to} & & c_k-p_{k} \geq \sup_{f \in \mathcal{F}}\left ( \sigma^2 + t\sqrt{s_f(p_k)} \right), \forall k \in [L], \\
    && & p_{k+1}-c_k \geq \sup_{f \in \mathcal{F}}\left ( t\sqrt{s_f(p_{k+1}) }- \sigma^2 \right), \forall k \in [L-1], \\
    & & &\frac{1}{L}\sum_{k =1}^{L}p_k = 1, p_k\geq 0, \forall k \in [L]. \\
  \end{aligned}
\end{equation}
% with a solution $t^*$ meaning that $$\tilde I_k(d_{R,k}) \geq
% 0.5(t^*)^2, \mbox{ and } \tilde I_k(d_{L,k}) \geq 0.5(t^*)^2,
% \forall k \in [L]$$ for all channel statistics in $\mathcal{F}$.
This problem is equivalent to finding the largest $t^*>0$ which gives
a feasible point in this formulation:
\begin{equation}
  \begin{aligned}
    \label{feasibilityproblem3}
    & \text{find} & & \{p_k\}_{k=1}^{L},\{c_k\}_{k=1}^{L-1} \\
    & \text{subject to} & & c_k-p_{k} \geq \sup_{f \in \mathcal{F}}\left ( \sigma^2 + t^*\sqrt{s_f(p_k)} \right), \forall k \in [L],\\
    && & p_{k+1}-c_k \geq \sup_{f \in \mathcal{F}}\left ( t^*\sqrt{s_f(p_{k+1}) }- \sigma^2 \right), \forall k \in [L-1],\\
    & & &\frac{1}{L}\sum_{k =1}^{L}p_k = 1, p_k\geq 0, \forall k \in [L]. \\
  \end{aligned}
\end{equation}
Solving the above feasibility problem can be done as follows: Fix a
small $t^*>0$ and choose $p_1^*=0$ and $d_{L,1}^*=\infty$ so that
$\bar I_1(d_{L,1}^*)=\infty$. Using $p_1^*>0$ would lead to a
sub-optimal solution since the transmitter has an average power
constraint and $s_f(p)$ is an increasing function of $p$ for every $f
\in \mathcal{F}$. Then, choose $c_1^*$ which satisfies $$c_1^*
=\sup_{f \in \mathcal{F}}\left(\sigma^2+
  t^*\sqrt{s_f(p_1^*)}\right)+p_1^*,$$ and as $p_2^*$, the minimum $p$
that satisfies
\begin{align}
  \label{eq:feasi}
  p-c_1^* -\sup_{f \in \mathcal{F}}\left
    (t^*\sqrt{s_f(p)}-\sigma^2\right ) \geq 0.
\end{align}
Note that for $0<t^*\leq \inf_{f \in
  \mathcal{F}}\frac{1}{\sqrt{\alpha_1}}$ there always exists a $p \geq
c_1^*-\inf_{f \in \mathcal{F}} \sigma^2$ that satisfies the above
equation. To see this, define the following auxiliary function $
w_f(p) = p-c_1^*-t^*\sqrt{s_f(p)}+\sigma^2,$ for which, for any fixed
$f \in \mathcal{F}$, it holds that $w_f(c_1^*-\sigma^2)<0$ and
$\lim\limits_{p\rightarrow \infty}\frac{w_f(p)}{p} =
1-t^*\sqrt{\alpha_1}>0.$

Note that choosing a higher value for $c_1^*$ would only make $p_2^*$
larger (or infinity) and thus, use more transmit power than necessary
(or make the problem infeasible). Using the same procedure we can
sequentially specify all $\{p_k^*\}_{k=1}^{L}$ and
$\{c_k^*\}_{k=1}^{L-1}$. Then, if $$\frac{1}{L}\sum_{k}^{L} p_k^* \leq
1,$$ the problem is feasible. However, if the average power constraint
is not satisfied, it is not possible to guarantee this error exponent
for all channels in $\mathcal{F}$, since in our construction, we pack
the decoding regions and constellation points as closely as
possible. To see this, if in the above construction we choose any
value $\tilde c_k > c_k^*$, then the corresponding $p$ which satisfies
\eqref{eq:feasi} would be larger than $p_{k+1}^*$ since $$p -\tilde
c_k -\sup_{f \in \mathcal{F}}\left (t^*\sqrt{s_f( p)}-\sigma^2\right )
\leq p - c_k^* -\sup_{f \in \mathcal{F}}\left (t^*\sqrt{s_f(
    p)}-\sigma^2\right ), \forall p>p_{k}^*+\tilde c_k.$$ \qed

\footnotesize \bibliography{references} \bibliographystyle{IEEEtran}
\end{document}